\documentclass[%
 preprint,
superscriptaddress,
nobibnotes,
 amsmath,amssymb,
 aps,
prb,
floatfix,
]{revtex4-2}

\usepackage[utf8]{inputenc}
\usepackage{graphicx}
\usepackage{dcolumn}% Align table columns on decimal point
\usepackage{bm}% bold math
\usepackage{xcolor}
\usepackage{microtype}
\usepackage[
    separate-uncertainty = true,
    locale=US,
    ]{siunitx}
\usepackage{cleveref}

\begin{document}

\title{Laser-induced real-space topology control of spin wave resonances} %magnon

\author{Tim Titze}
\affiliation{I.\,Physikalisches Institut, Universit\"at G\"ottingen, 37077 G\"ottingen, Germany\looseness=-1}
\author{Sabri Koraltan}
\affiliation{Physics of Functional Materials, Faculty of Physics, University of Vienna, Vienna, Austria\looseness=-1}
\affiliation{Vienna Doctoral School in Physics, University of Vienna, Vienna, Austria}
\author{Timo Schmidt}
\affiliation{Institute of Physics, University of Augsburg, 86135 Augsburg, Germany\looseness=-1}
\author{Marcel Möller}
\affiliation{Max Planck Institute for Multidisciplinary Sciences, G\"ottingen, Germany\looseness=-1}
\affiliation{IV.\,Physikalisches Institut, Universit\"at G\"ottingen, 37077 G\"ottingen, Germany}
\author{Florian Bruckner}
\affiliation{Physics of Functional Materials, Faculty of Physics, University of Vienna, Vienna, Austria\looseness=-1}
\author{Claas Abert}
\affiliation{Physics of Functional Materials, Faculty of Physics, University of Vienna, Vienna, Austria\looseness=-1}
\affiliation{Research Platform MMM Mathematics-Magnetism-Materials, University of Vienna, Vienna, Austria}
\author{Dieter Suess}
\affiliation{Physics of Functional Materials, Faculty of Physics, University of Vienna, Vienna, Austria\looseness=-1}
\affiliation{Research Platform MMM Mathematics-Magnetism-Materials, University of Vienna, Vienna, Austria}
\author{Claus Ropers}
\affiliation{Max Planck Institute for Multidisciplinary Sciences, G\"ottingen, Germany\looseness=-1}
\affiliation{IV.\,Physikalisches Institut, Universit\"at G\"ottingen, 37077 G\"ottingen, Germany}
\affiliation{International Center for Advanced Studies of Energy Conversion (ICASEC), Universit\"at G\"ottingen, 37077 G\"ottingen, Germany}
\author{Daniel Steil}
\affiliation{I.\,Physikalisches Institut, Universit\"at G\"ottingen, 37077 G\"ottingen, Germany\looseness=-1}
\author{Manfred Albrecht}
\affiliation{Institute of Physics, University of Augsburg, 86135 Augsburg, Germany\looseness=-1}
\author{Stefan Mathias}
\affiliation{I.\,Physikalisches Institut, Universit\"at G\"ottingen, 37077 G\"ottingen, Germany\looseness=-1}
\affiliation{International Center for Advanced Studies of Energy Conversion (ICASEC), Universit\"at G\"ottingen, 37077 G\"ottingen, Germany}

\begin{abstract}
   Femtosecond laser excitation of materials that exhibit magnetic spin textures promises advanced magnetic control via the generation of ultrafast and non-equilibrium spin dynamics. We explore such possibilities in ferrimagnetic [Fe(0.35 nm)/Gd(0.40 nm)]$_{160}$ multilayers, which host a rich diversity of magnetic textures from stripe domains at low magnetic fields, a dense bubble/skyrmion lattice at intermediate fields, and a single domain state for high magnetic fields. Using femtosecond magneto-optics, we observe distinct coherent spin wave dynamics in response to a weak laser excitation allowing us to unambiguously identify the different magnetic spin textures. Moreover, employing strong laser excitation we show that we achieve versatile control of the coherent spin dynamics via non-equilibrium and ultrafast transformation of magnetic spin textures by both creating and annihilating bubbles/skyrmions. We corroborate our findings by micromagnetic simulations and by Lorentz transmission electron microscopy before and after laser exposure.
\end{abstract}

\maketitle
\section{Introduction}

Magnetic spin textures are expected to be an important building block for future spintronic- and magnonic-based memory and logic devices, and even for unconventional computing techniques such as neuromorphic computing~\cite{Finocchio2016, Fert2017, Prychynenko2018, Song2020, Zhang2020, Chumak2022}. Especially, localized magnetic solitons, known as skyrmions, have attracted significant attention to be utilized in spin texture based devices~\cite{Bogdanov1994, Roessler2006, Yu2010}. Potential applications of magnetic textures include the engineering of spin wave dispersions in magnonic crystals, using magnetic textures as spin wave phase shifters in logic devices, and employing them as nanoscale spin wave emitters to name a few~\cite{Lenk2011, Yu2021, Petti2022}. Obviously, the ability to actively control such devices by non-destructive and fast means is highly desirable from an application point of view. Ultrafast optical manipulation of spin textures is a potential avenue to achieve these goals. However, optical spectroscopy studies in spin-textured materials systems have so far either focused on precessional spin dynamics present in low-temperature skyrmionic phases~\cite{Ogawa2015, Padmanabhan2019, Sekiguchi2022, Kalin2022}, or on statically observed laser-induced transformations of spin textures~\cite{Eggebrecht2017, Je2018, Buettner2021, Khela2023}. Here we show that we can achieve ultrafast optical control of spin textures at room temperature and thus laser-induced tuning of spin wave resonances to specific frequency ranges. Our work opens new avenues for future actively controlled magnetic spin texture-based devices.

\section{Results}
\subsection{Static magnetic properties of [Fe(\SI{0.35}{\nano \meter})/Gd(\SI{0.40}{\nano \meter})]$_{160}$ thin films}

\begin{figure}[htb]
     \centering
     \includegraphics[width=\columnwidth]{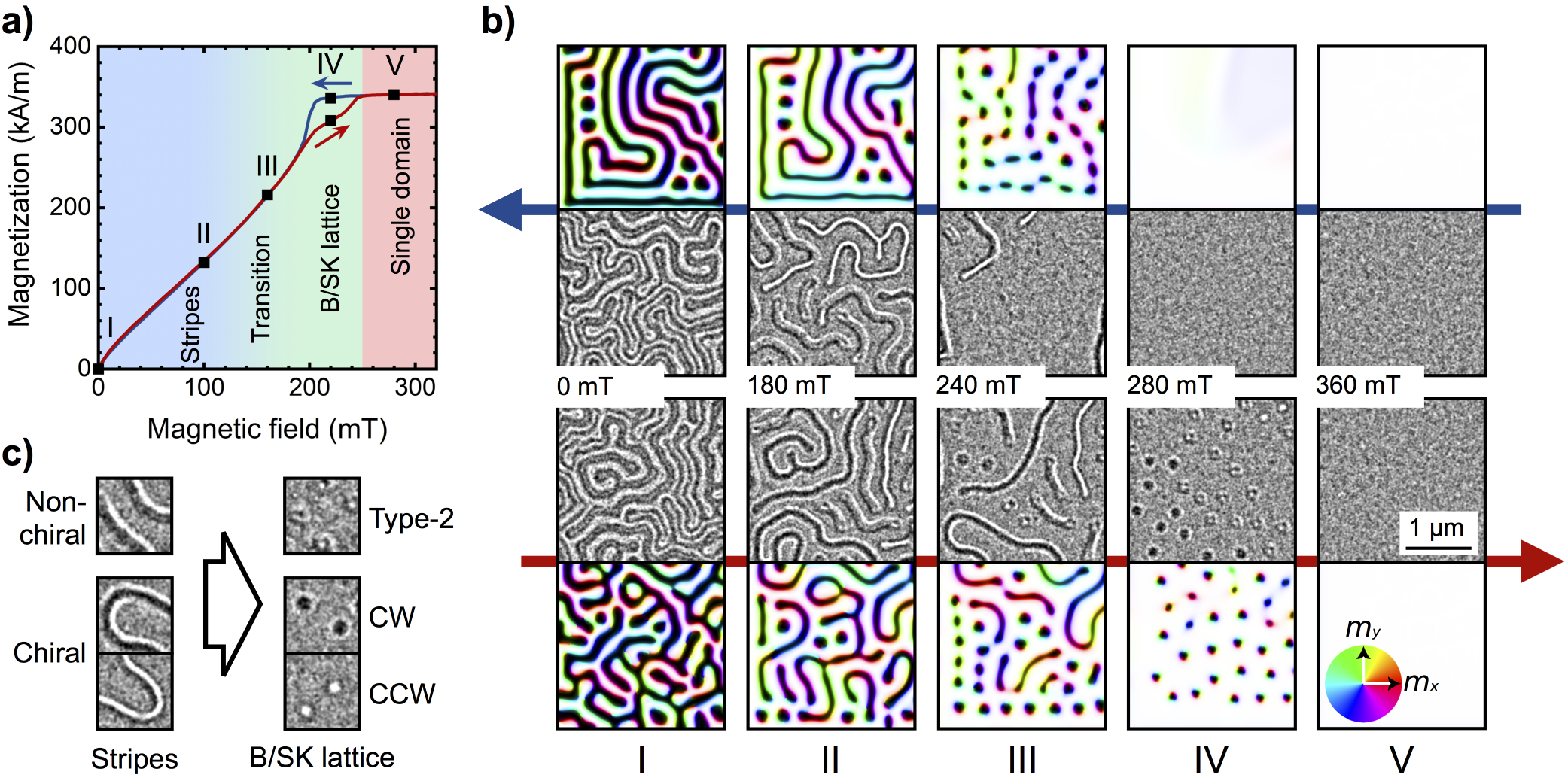}
     \caption{\textbf{Static magnetic sample properties:} \textbf{a)} Out-of-plane $M$-$H$ hysteresis loop of the [Fe(\SI{0.35}{\nano \meter})/Gd(\SI{0.40}{\nano \meter})]$_{160}$ thin film at room temperature. Solid lines correspond to $M$-$H$ data for increasing (red) and decreasing (blue) field, black squares mark assignments of LTEM images (I)--(V) in b) to magnetic field regions in the $M$-$H$ loop. The coloration of magnetic field regimes represent the underlying spin textures. \textbf{b)} LTEM images (grayscale) and micromagnetic simulations (color) of the magnetization for an upsweep (bottom, red arrow) and downsweep (top, blue arrow) of the magnetic field reveal the spin texture corresponding to different magnetic fields (I)-(V) in the hysteresis loop. The direction of $M$ in the micromagnetic simulation data is given by the color code. \textbf{c)} Legend of different spin objects indicating the field-induced transformations possible for a magnetic field upsweep with non-chiral and chiral domains walls, type-2 bubbles and clockwise (CW) and counter-clockwise (CCW) skyrmions.
     }
     \label{fig:LTEM}
\end{figure}

The studied sample system is a multilayer stack of Fe and Gd with composition [Fe(0.35 nm)/Gd(0.40 nm)]$_{160}$ deposited on a thermally-oxidized Si(100) substrate. Static magnetic characterization of this sample was performed using Superconducting Quantum Interference Device - Vibrating Sample Magnetometry (SQUID-VSM). Samples deposited on thin commercial Si$_3$N$_4$ membranes were used for Lorentz Transmission Electron Microscopy (LTEM) studies to image the static magnetic texture in applied out-of-plane (oop) magnetic fields.

Figure~\ref{fig:LTEM}a shows the $M$-$H$ hysteresis loop from SQUID-VSM for positive magnetic fields and Fig.~\ref{fig:LTEM}b displays the corresponding LTEM images (gray-scaled middle panels) for representative field values together with micromagnetic simulations of the field-dependent magnetic texture (color-scaled top and bottom panels). Without applied field (Figs.~\ref{fig:LTEM}a and~\ref{fig:LTEM}b, left), the sample is in a demagnetized state, and the magnetic texture consists solely of oop stripe domains with Bloch domain walls (region I). We find two different types of domain walls, similar to previous studies~\cite{Montoya2017a, Heigl2021}: Narrow stripes with domain walls that do not exhibit any chirality, and broader stripes with twice the periodicity featuring the same chirality. Applying an increasing oop magnetic field (Fig.~\ref{fig:LTEM}a, red part of $M$-$H$ loop), the oop magnetization of the sample initially increases linearly with the applied magnetic field. In LTEM, this process is visible as a decrease of the density of domain walls (Fig.~\ref{fig:LTEM}b, II, bottom images) and the nucleation of cylindrical spin objects from collapsed narrow stripe domains (so called type-2 magnetic bubbles). For higher magnetic fields (region III), dipolar Bloch-type skyrmions start to nucleate from the broad stripes. In region IV, in a field range of $\mu_0 H\approx \SI{190}{}-\SI{250}{\milli \tesla}$, a bubble and skyrmion (B/SK) lattice with characteristic spacing and groups of spin objects of the same type is formed. Close to saturation, the bubbles vanish and only skyrmions remain (see SI Fig.~\ref{fig:BubblesSkyrmions}). For even higher applied fields (region V) magnetic saturation is achieved. Decreasing the magnetic field from saturation does not lead to the appearance of a B/SK lattice in region IV, as reflected in the LTEM images (Fig.~\ref{fig:LTEM}b, IV, top) and $M$-$H$ loop (Fig.~\ref{fig:LTEM}a, blue line) and corroborated by the micromagnetic simulation. The measured oop magnetization in this region is larger than for the upsweep of the magnetic field, as no cylindrical spin objects reducing the oop magnetization exist. The small hysteresis in this field region in the $M$-$H$ loop is thus indicative of the disordered B/SK lattice. For further decreasing field strengths (regions III to I), the evolution of the magnetic texture is again comparable to the upsweep, except that, at least in region III, the initial domain wall density is lower in the decreasing field case. Please note that small differences observed in the magnetic sample properties between SQUID-VSM and LTEM measurements, i.e., higher magnetic field values for the same spin texture in LTEM, arise from the fact that two different substrates are used for the sample preparation resulting in different growth conditions of the films. Micromagnetic simulations match the magnetic field values of the corresponding SQUID-VSM data and are discussed further below.

\subsection{Coherent spin dynamics in response to weak optical excitations}

After the thorough characterization of the ground state magnetic spin textures, we now focus on the magnetic response of the sample to an ultrashort optical excitation. In particular, we seek correlations between the temporal response and the different spatial textures in Fig.~\ref{fig:LTEM}. For these measurements, we use a comparably weak perturbation (fluence $F=\SI{300}{\micro \joule \per \centi \meter \squared}$, pulse duration $\tau_p<40$\,fs) leading to a total demagnetization of the sample by only \SI{3.5}{\percent}. Magnetization dynamics $M(t)$ were detected by changes in the light polarization upon reflection of the sample using the magneto-optical Kerr effect (MOKE) in polar geometry with the magnetic field applied in oop direction. For details see the methods section.

\begin{figure}[htb]
     \centering
     \includegraphics[width=0.7\columnwidth]{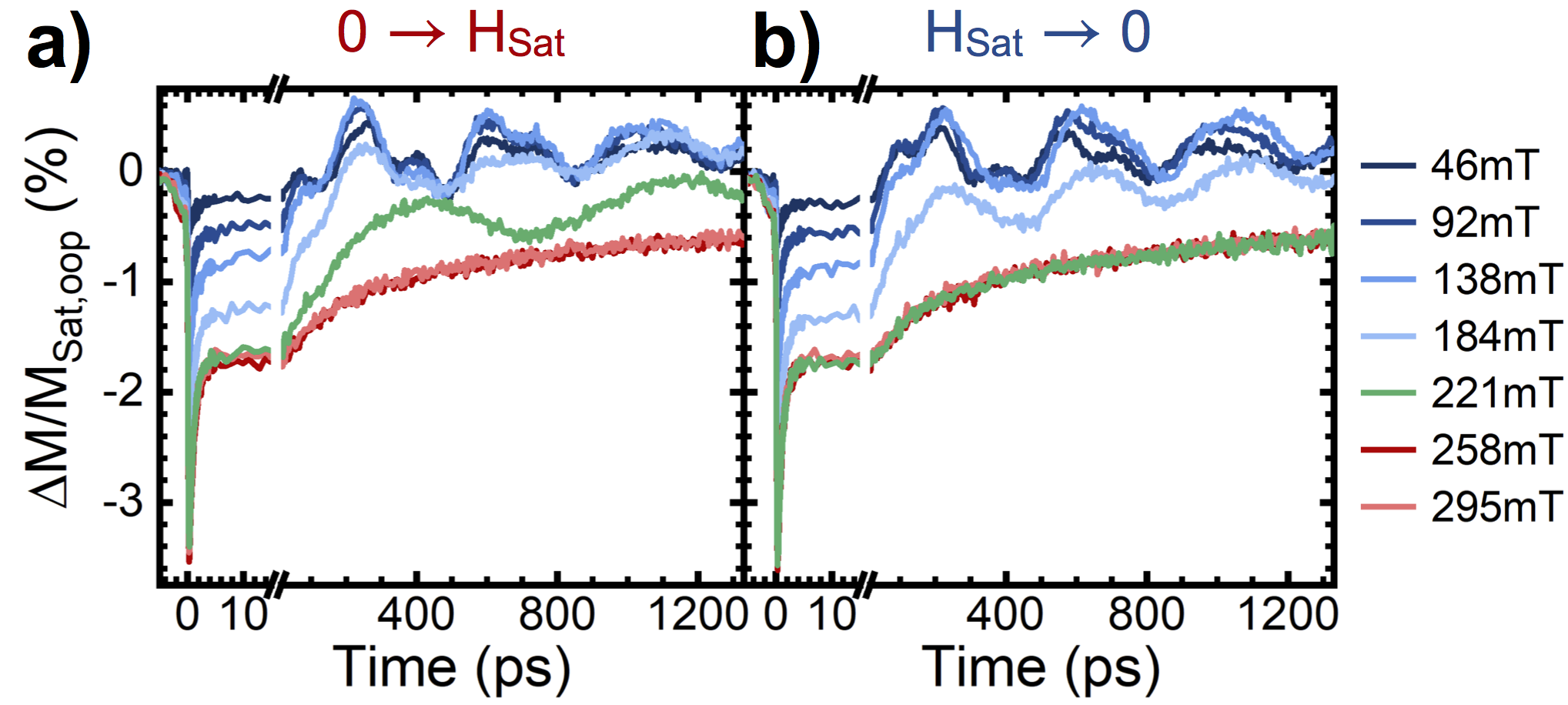}
     \caption{\textbf{Out-of-plane magnetization dynamics for a weak perturbation in dependence of the applied oop magnetic field.} Magnetization dynamics for \textbf{a)} upsweep and \textbf{b)} downsweep of the magnetic field. The colors of the curves represent the regions of different spin texture as given in Fig.~\ref{fig:LTEM}a: Stripe domains (blue), B/SK lattice (green), and magnetic saturation (red). Data is normalized to the Kerr signal in magnetic saturation in oop direction.}
     \label{fig:overview20mW}
\end{figure}

Figure~\ref{fig:overview20mW} depicts the magnetization dynamics in response to the optical excitation and in dependence of the applied oop magnetic field for both an upsweep (Fig.~\ref{fig:overview20mW}a) and a downsweep of the magnetic field (Fig.~\ref{fig:overview20mW}b). Four processes can be identified: Immediately after photoexcitation, the magnetization is quenched on a timescale of only \SI{300}{\femto \second}. This ultrafast demagnetization process is followed by a fast remagnetization on a few picosecond timescale. During the remagnetization, about half of the initial magnetization is recovered, and the magnetization stays constant thereafter for up to $\approx$\SI{40}{\pico \second} depending on the applied magnetic field. Finally, the magnetization recovers on a \SI{}{\nano \second}-timescale by thermal transport out of the film. Most notably, in addition to this well-known and incoherent magnetic response~\cite{Beaurepaire1996, Kirilyuk2010}, a coherent oscillatory signal component is clearly visible in the stripe domain phase (blue transients) and the B/SK lattice phase for the upsweep of the magnetic field (green transient).

\begin{figure}[htb]
     \centering
     \includegraphics[width=\columnwidth]{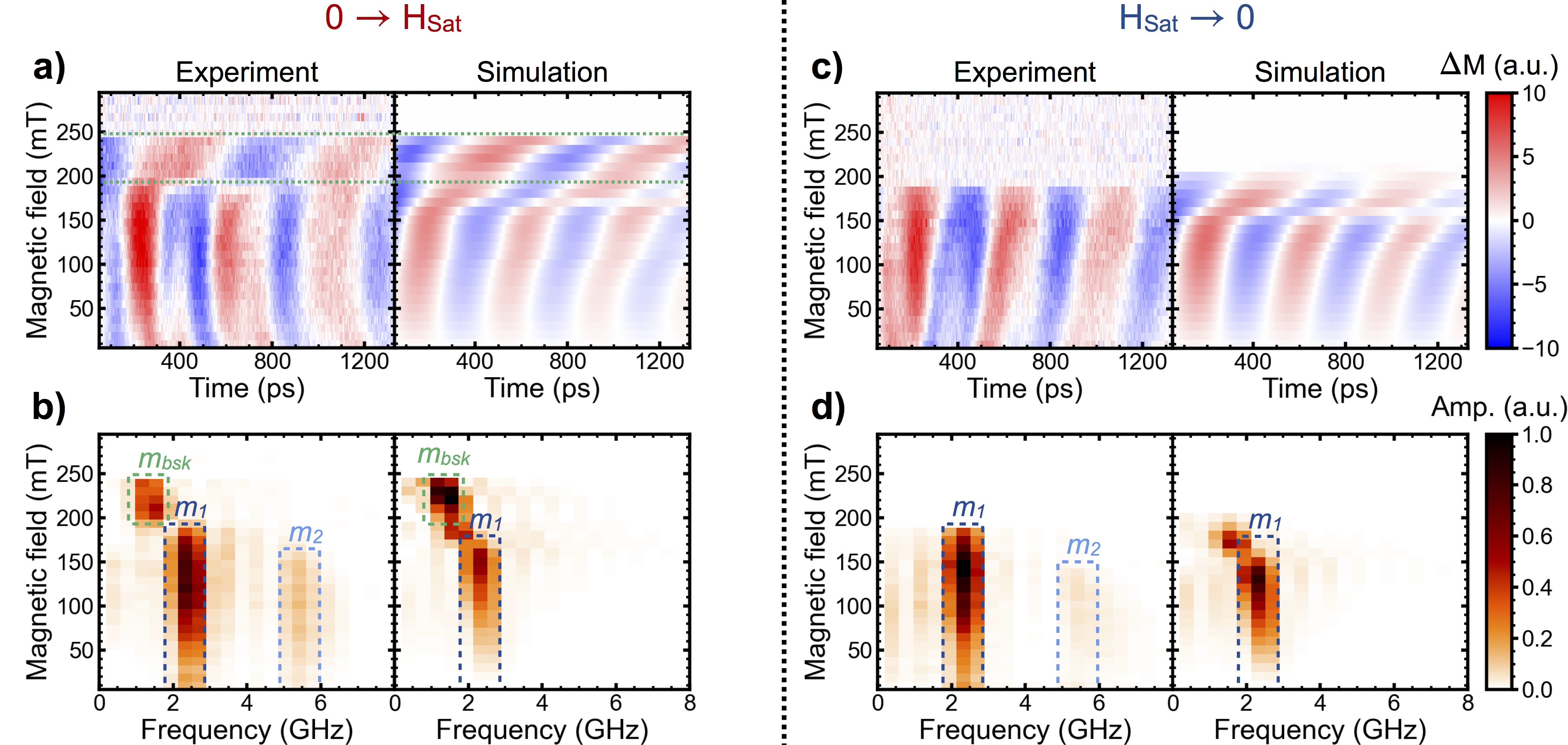}
     \caption{\textbf{Coherent magnetization dynamics in the experiment and micromagnetic simulation in dependence of the applied oop magnetic field.} Coherent magnetization dynamics for \textbf{a)} upsweep and \textbf{c)} downsweep of the magnetic field with the corresponding Fourier spectra \textbf{b)} and \textbf{d)}, respectively. In case of the upsweep \textbf{a)}, \textbf{b)} a unique spin wave mode is observed for $\mu_0 H=\SI{190}{}-\SI{240}{\milli \tesla}$.}
     \label{fig:oscTheoExp}
\end{figure}
The main question now is to what extent this coherent response is indicative of the specific spin texture in the Fe/Gd sample. To answer this question, we performed micromagnetic simulations of the spin dynamics after a magnetization quench of less than \SI{10}{\percent} (see Methods). Figure~\ref{fig:oscTheoExp} depicts a full comparison of the experimentally found and simulated coherent oscillatory signal components after subtraction of the incoherent background for both the upsweep (Fig.\,\ref{fig:oscTheoExp}a) and downsweep (Fig.\,\ref{fig:oscTheoExp}c) of the magnetic field. The corresponding Fourier transforms for experiment and micromagnetic simulation are shown in Fig.~\ref{fig:oscTheoExp}b~and~Fig.~\ref{fig:oscTheoExp}d, respectively.

Evidently, experiment and simulation match well and show a clear correspondence of the underlying spin textures with the observed coherent spin dynamics. The Fourier spectra for magnetic fields of $\mu_0 H = 0 - 190$\,mT, i.e., the stripe domain state, show a strong oscillatory mode $m_1$ with $f_1\approx 2.3$\,GHz, which is identified as a breathing mode of stripe domains in the micromagnetic simulation. A much weaker high-frequency component $m_2$ with $f_2\approx 5.4$\,GHz is additionally observed in the experiment, however, its origin needs to be further investigated. For magnetic fields of $\mu_0 H \approx 190 - 240$\,mT the disordered B/SK lattice is present in the upsweep of the magnetic field only. Correspondingly, experiment and simulation show only in this case an oscillatory mode $m_{bsk}$ with $f_{bsk}\approx 1.4$\,GHz. This frequency is identified as a breathing mode of bubbles and skyrmions~\cite{Mochizuki2012} in the micromagnetic simulation in very good agreement with previous studies on skyrmion dynamics~\cite{Sekiguchi2022,Ogawa2015,Onose2012}. We note that micromagnetic simulations indeed predict a slightly lower frequency for magnetic bubbles compared to skyrmions, however, we cannot resolve this difference with our experimental frequency resolution. Last, in saturation, no oscillatory spin dynamics are observed in experiment and simulation.

While the agreement between experiment and simulation is very good in general, we note that small differences do exist in the magnetic field range of $\mu_0 H \approx 150 - 200$\,mT for both field up- and downsweep. Furthermore, there is an apparent phase shift in the oscillatory dynamics with magnetic field in both experiment and simulation. For details on these observations we refer to the Supplementary Information Section~\ref{supp_pert}.

\subsection{Magnetic phase transformation in response to a strong optical excitation}
\begin{figure}[htb]
     \centering
     \includegraphics[width=0.7\columnwidth]{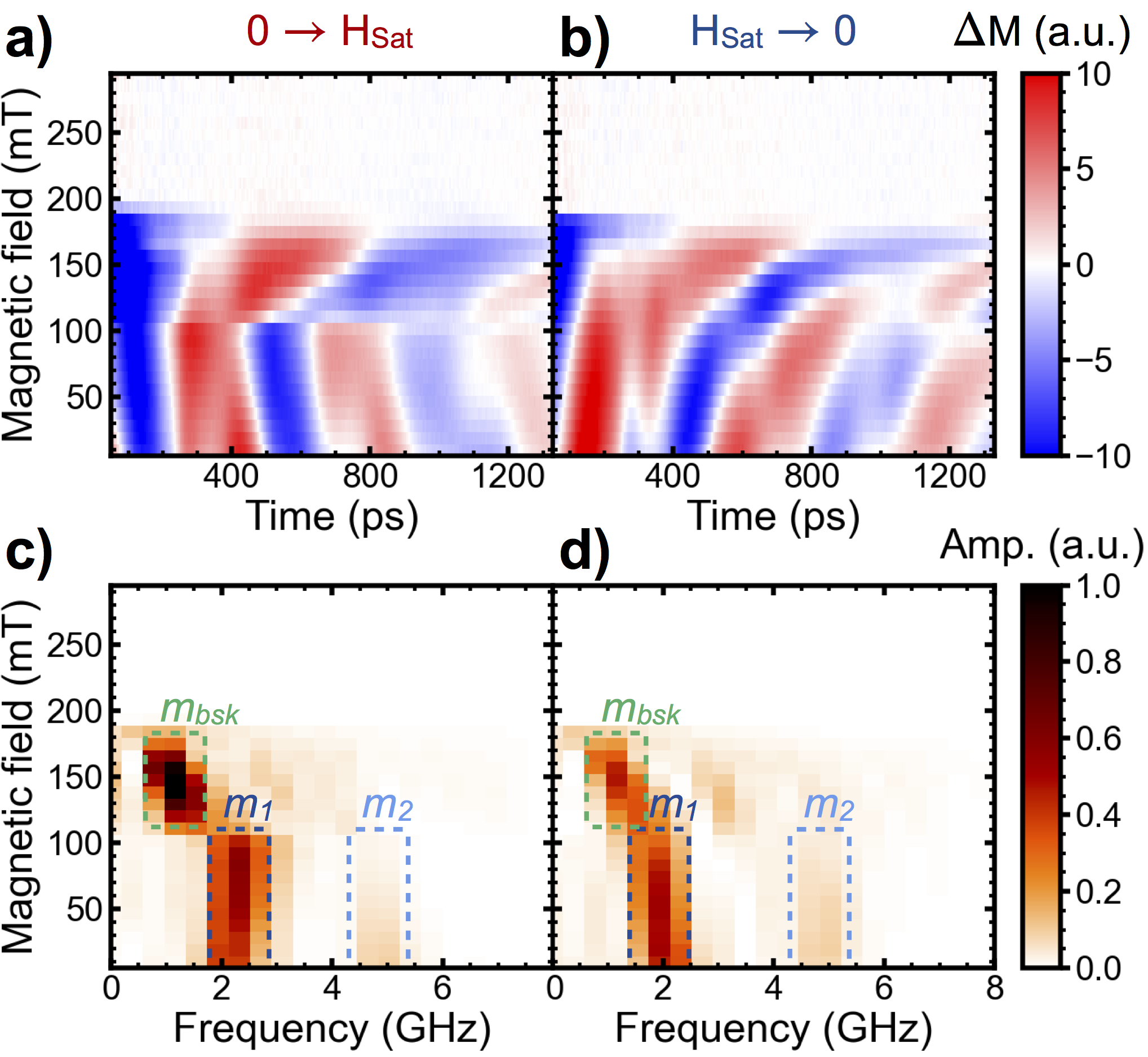}
     \caption{\textbf{High-fluence coherent magnetization dynamics in time-resolved MOKE in dependence of the applied oop magnetic field.} High fluence coherent magnetization dynamics for \textbf{a)} upsweep and \textbf{b)} downsweep  of the magnetic field with the corresponding Fourier spectrum \textbf{c)} and \textbf{d)}, respectively. For both directions of magnetic field sweep a single spin wave mode is observed within $\mu_0 H=\SI{100}{}-\SI{190}{\milli \tesla}$.}
     \label{fig:oscpart200mW}
\end{figure}

Having shown a clear correlation between spin texture and spin dynamics, we now focus on the response of the spin texture to a strong non-equilibrium excitation of the spin system. We repeat the experiments from above for a strong laser excitation of $F=\SI{3}{\milli \joule \per \centi \meter \squared}$, i.e., an order of magnitude larger fluence than before, and we find significant changes in the observed dynamics. A much stronger incoherent demagnetization of about 45\% occurs, i.e., the initial magnetic spin texture is strongly disturbed (see SI Figs.~\ref{fig:overview200mW}a~and~b). For a detailed investigation, we again subtracted the incoherent background (Fig.~\ref{fig:oscpart200mW}a~and~b) and performed a Fourier transform (Fig.~\ref{fig:oscpart200mW}c~and~d).

We now observe the stripe domain phase mode $m_1$ in a magnetic field range $\mu_0 H=0-\SI{100}{\milli \tesla}$ with slightly lower frequency than before. A second, higher-frequency mode $m_2$ is weakly present at $f_2\approx\SI{4.8}{\giga \hertz}$. The mode previously identified to be B/SK breathing now appears around $f_{bsk}\approx\SI{1.2}{\giga \hertz}$ in the magnetic field range $\mu_0 H=100-\SI{190}{\milli \tesla}$. In contrast to the weak perturbation limit, the mode now exhibits a clear frequency dependence, shifting from  $f\approx\SI{1.4}{\giga \hertz}$ to $f\approx\SI{1.0}{\giga \hertz}$ from low to high magnetic fields. Above $\mu_0 H=\SI{190}{\milli \tesla}$ no oscillatory behavior is observed anymore. Most surprisingly, however, we now observe the $m_{bsk}$ mode even in the downsweep of the magnetic field (Fig.~\ref{fig:oscpart200mW}b~and~d), where it was previously absent. This observation suggests that the strong laser excitation itself might induce a B/SK lattice phase in the material. Indeed, skyrmion nucleation from a saturated state by strong laser excitation was recently theoretically predicted~\cite{Koshibae2014} and experimentally observed~\cite{Je2018, Buettner2021}, which could explain our findings. However, our micromagnetic simulations strongly suggest that a nucleation of skyrmions from saturation does not appear to be possible for the Fe/Gd multilayers studied here.

To understand the observed behavior better, we performed LTEM measurements before and after femtosecond laser excitation. These measurements indeed allow us to identify the changes in dynamics from Fig.~\ref{fig:oscTheoExp} to Fig.~\ref{fig:oscpart200mW} as a laser-induced transformation of the underlying spin texture. We will show in the following that this transformation is not trivial, and not dominantly given by a simple static laser-induced temperature shift of phase stability ranges~\cite{Montoya2017a} together with laser-induced skyrmion nucleation from magnetic saturation~\cite{Koshibae2014,Je2018, Buettner2021}.

\begin{figure}[h!]
     \centering
     \includegraphics[width=\columnwidth]{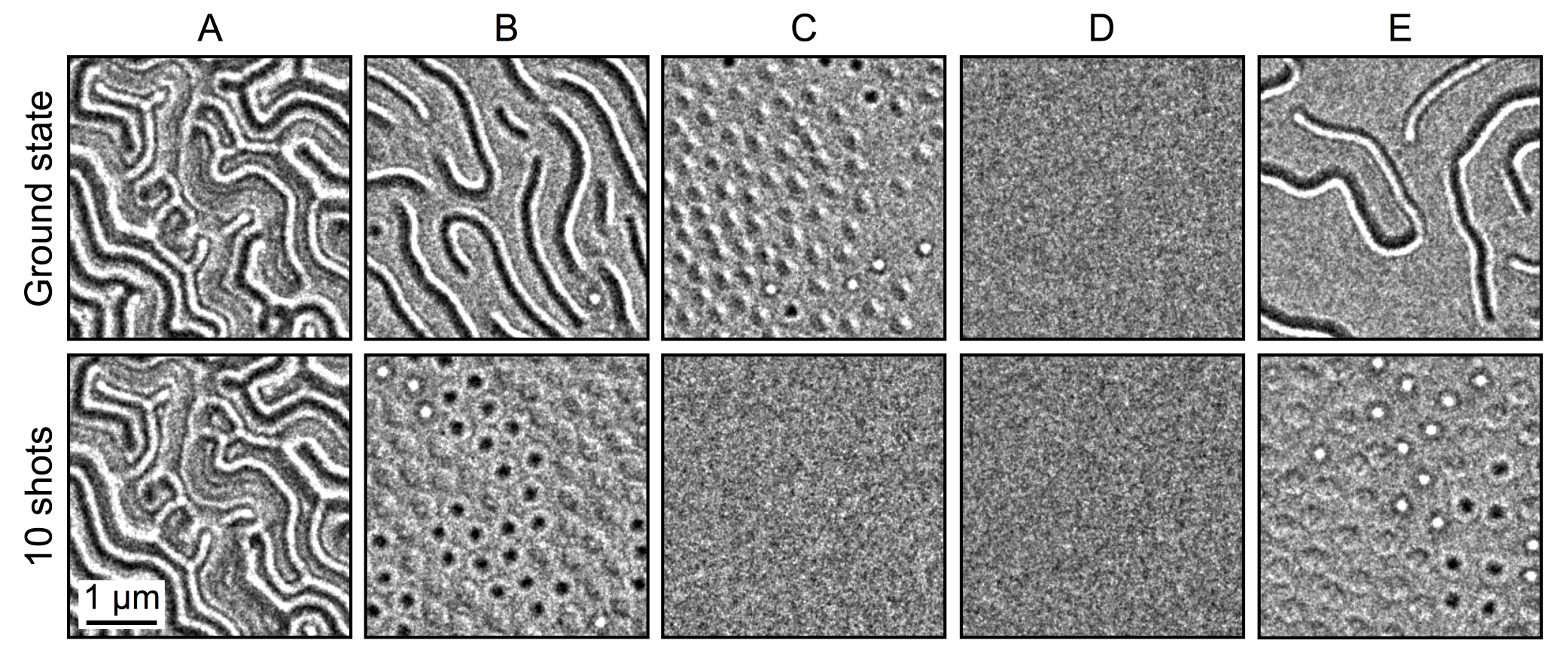}
     \caption{\textbf{Laser-induced transformation of magnetic textures in LTEM.} LTEM experimental data showing the magnetic texture before and after strong laser excitation of $F=\SI{6.4}{\milli \joule \per \centi \meter \squared}$. \textbf{A)}~Unmodified stripe domain phase for low magnetic fields; \textbf{B)} transformation of a mixed phase into a B/SK lattice phase;  \textbf{C)} annihilation of bubbles and skyrmions;  \textbf{D)} unmodified saturated magnetic state; \textbf{E)} transformation of a stripe state in the downsweep of the magnetic field into a mixed state (see SI Fig.~\ref{fig:ShotsLTEMScE}) and then further into a B/SK lattice state.}
     \label{fig:ShotsLTEM}
\end{figure}

Upon ultrafast and strong laser excitation, we distinguish five scenarios depending on the field-dependent magnetic ground state of the system depicted in Fig.~\ref{fig:ShotsLTEM} starting with the upsweep of the magnetic field:
A) For low magnetic fields, a pristine stripe domain phase remains nearly unchanged after laser excitation. B) A former mixed stripe domain and B/SK state is turned into a pure B/SK lattice phase, as also observed in the simulation, see SI Fig.~\ref{fig:SkNucSim}. C) Starting within the B/SK lattice phase, bubbles and skyrmions are annihilated by the laser excitation, i.e., the system is driven into a fully saturated state even by a single laser shot. Thereafter, the new saturated magnetic state is not modified by further laser excitation.  D) From saturation, we did not observe any laser-induced creation of skyrmions without irreversibly changing the sample properties, which is in contrast to literature~\cite{Koshibae2014,Je2018,Buettner2021}. However, this observation concurs with the micromagnetic simulations, where it was also not possible to nucleate skyrmions from a saturated spin state. Note that close to saturation in the downsweep of the magnetic field, a gradual transformation of the system starting from a single isolated stripe domain appears possible (see SI Fig.~\ref{fig:SkfromSt}). E) For the downsweep, a plain stripe domain phase with low stripe density is transformed into a mixed phase. Here, all stripe domains convert into cylindrical spin textures after exposure to a single laser pulse and new stripe domains nucleate in regions that are initially saturated (see SI Fig.~\ref{fig:ShotsLTEMScE}). Upon subsequent laser excitation the newly generated domains transform into B/SKs. Hence, using ultrafast and strong laser excitation, a pure B/SK lattice phase can be created. A schematic phase diagram of the magnetic texture depending on magnetic field and the number of incident laser pulses can be found in SI Fig.~\ref{fig:Phasediagram}.

In comparison to literature, for a Co/Pt multilayer sample, Gerlinger and coworkers~\cite{Gerlinger2021} observed laser-induced skyrmion nucleation from both i) magnetic saturation and ii) the stripe domain phase. Furthermore, they observed iii) an annihilation of skyrmions and stripe domains close to saturation. In our study, we obviously find similar evidence for processes ii) and iii), but not for i), the laser-induced nucleation from saturation.

However, taking the predictions from micromagnetic simulations and the observations from the LTEM measurements  into account, we can now understand the changes in the coherent oscillatory component that come along with the strong excitation in Fig.~\ref{fig:oscpart200mW}: In the upsweep of the magnetic field, the former B/SK lattice phase completely vanishes before the actual data recording starts due to continuous excitation by laser pulses in the time-resolved MOKE experiment. Hence, we observe no oscillatory component, but a signal corresponding to magnetic saturation in the magnetic field range $\SI{190}{}<\mu_0 H<\SI{250}{\milli \tesla}$ (scenario C, Fig.~\ref{fig:ShotsLTEM}). For lower magnetic fields, the initially mixed stripe domain and B/SK phase is laser-transformed into a pure B/SK lattice phase, leading to the apparent shift of the $m_{bsk}$ mode to lower frequency at high fields, $\SI{150}{}<\mu_0 H<\SI{190}{\milli \tesla}$ (scenario B, Fig.~\ref{fig:ShotsLTEM}). The slightly higher frequency observed in Fig.~\ref{fig:oscpart200mW}c for magnetic fields of about $\SI{110}{}<\mu_0 H<\SI{150}{\milli \tesla}$ originates from a mixed phase that forms upon laser excitation out of a pristine stripe domain phase exhibiting none or only very few skyrmions, respectively. This means that no final pure B/SK lattice phase is created within this field range (see also SI Fig.~\ref{fig:Superposition}). In the downsweep of the magnetic field, for high fields ($\SI{150}{}<\mu_0 H<\SI{190}{\milli \tesla}$) a pure B/SK lattice phase is formed out of a pure stripe phase (scenario E, Fig.~\ref{fig:ShotsLTEM}), and, for lower fields ($\SI{100}{}<\mu_0 H<\SI{150}{\milli \tesla}$), a mixed phase is the final state. All these mechanisms lead to the generally more equal oscillatory behavior observed for the strong perturbation case for magnetic field up- and downsweep. However, the absence of an initially mixed phase in the downsweep hinders the generation of a B/SK lattice phase. For equal magnetic fields we observe less bubbles and skyrmions being created in the downsweep of the magnetic field resulting in a dominance of the stripe's oscillatory component in the generated mixed phase. Finally, the generally lower frequencies of all modes can most likely be attributed to the transient laser-induced change of macroscopic sample properties, e.g., sample temperature, magnetization and anisotropies~\cite{Montoya2017a}.

Closing the discussion, three further points are noteworthy: First, an asymmetry in the system response exists for the upsweep and downsweep of the magnetic field in the field regions where no B/SKs are observed in static studies. This asymmetry increases with a stronger perturbation. This is evident from the different shifts of magnetic oscillations for the upsweep and downsweep of the magnetic field in Fig.~\ref{fig:oscTheoExp}a~and~c as well as Fig.~\ref{fig:oscpart200mW}a~and~b, which even shows a frequency shift between the up and down field sweeps (cf. Fig.~\ref{fig:oscpart200mW}c~and~d). However, SQUID-VSM and LTEM data from Fig.~\ref{fig:LTEM} do not suggest any specific macroscopic difference in the low-field range for both field sweep directions, except possibly in domain wall density. Currently, we do not have an explanation for this behavior, but aim to investigate this further in the future. Second, in contrast to Refs.~\cite{Je2018, Buettner2021, Gerlinger2021}, skyrmion creation in our case is only possible if there is an initially existing chiral spin texture. Even further, the nucleation of skyrmions does not only have a lower but also an upper fluence threshold. Above a certain fluence, a mixed phase is formed in the region of maximum excitation that is surrounded by a pure skyrmion phase due to the Gaussian intensity distribution of the laser pulse, see Fig.~\ref{fig:Fluence} in the SI. Last, we want to note that no influence of spin texture on the incoherent magnetization dynamics was observed, neither on the sub-ps timescale nor during the remagnetization.

\section{Conclusion}
In summary, we studied the response of the spin-textured material Fe/Gd to weak and strong ultrafast laser excitation. In the weak excitation limit, unique coherent spin dynamics evolve that allowed us to directly identify the magnetic stripe domain phase, the bubble/skyrmion lattice phase, as well as magnetic saturation. In the strong excitation limit, we observed significant changes in the dynamic magnetic phase diagram, which we attributed to a laser-induced transformation of the spin texture as verified by micromagnetics and LTEM measurements. Indeed, from the stripe domain state, we created both a bubble/skyrmion lattice and a mixed phase of bubbles, skyrmions and stripe domains, depending on the external magnetic field. These results highlight that ultrafast optical control of spin wave resonances is possible by optical transformation of the spin textures in the material. Moreover, these rich spin physics are observed at room temperature and at quite moderate magnetic fields and laser fluence, making the Fe/Gd multilayer system an attractive model system for fundamental studies of spin texture-based physics, as well as potential application in spin-electronics in the future.

\section{Methods}
\subsection{Sample preparation and characterization}

[Fe(0.35 nm)/Gd(0.40 nm)]$_{160}$ multilayer samples were prepared on both thermally-oxidized Si(100) substrates, as well as on 30-nm thick Si$_3$N$_4$ membranes required for LTEM imaging. The films were deposited at room temperature by dc magnetron sputtering in one run. Nevertheless, slightly varying magnetic properties were obtained due to different growth conditions during deposition on the two different substrates. The sputter process was carried out using an Ar working pressure of 3.5 µbar in an ultra-high vacuum chamber. For all samples, a 5-nm-thick Pt seed layer and a 5-nm-thick Si$_3$N$_4$ capping layer were used to protect the films from oxidation.
The thickness of the layers was controlled by a calibrated quartz balance during layer deposition. \\
The magnetic properties of the samples were analyzed by Superconducting Quantum Interference Device-Vibrating Sample Magnetometry (SQUID-VSM). $M$-$H$ hysteresis loops were measured both in out-of-plane  and in-plane configuration at room temperature. The magnetic spin textures were imaged by Lorentz Transmission Electron Microscopy (LTEM) at room temperature using a JEOL NEOARM-200F system operated at \SI{200}{\kilo \electronvolt} beam energy in the Fresnel mode with an underfocus of 2 mm. Images are recorded with a Gatan OneView camera in external out-of-plane magnetic fields.

\subsection{Time-resolved magneto-optical Kerr spectroscopy (TR-MOKE)}
Transient reflectivity $\Delta R(t)$ and magnetization dynamics $\Delta M(t)$ were measured simultaneously using a bichromatic pump-probe setup. Sub-\SI{}{\pico \second} laser pulses are generated by a fiber amplifier system operating at a repetition rate of \SI{50}{\kilo \hertz}. A pulse duration of less than \SI{40}{\femto \second} is achieved at the sample position via spectral broadening in a gas-filled hollow core fiber and subsequent compression with chirped mirrors \cite{moller_ultrafast_2021}. In the setup, the fundamental light pulses with central wavelength of \SI{1030}{\nano \meter} and their second harmonic at \SI{515}{\nano \meter} are used as pump and probe pulses, respectively. External out-of-plane magnetic fields up to $\mu_0 H=\SI{0,7}{\tesla}$ can be applied by mounting the sample in-between the poles of a variable-gap electromagnet. The data is measured by a balanced bridge detector and collected by a \SI{250}{\mega \hertz} digitizer card enabling single pulse detection. Using a mechanical chopper to improve signal quality, both the pumped and the unpumped signals can be recorded, allowing for maximum scope of data post processing. Data was acquired in polar Kerr geometry, where the magnetic field was applied in oop direction. All $M(t)$ data shown were recorded as the difference of $M(t)$ traces for two opposite magnetic fields to remove nonmagnetic signal components, taking care to set the correct spin texture by approaching the current magnetic field either from below or above.

\subsection{Lorentz Transmission Electron Microscopy with \textit{in-situ} optical excitation}
To obtain a better understanding of the spin dynamics observed in TR-MOKE in the strong excitation limit, we studied a sample on a thin commercial Si$_3$N$_4$ membrane by LTEM at the Göttingen Ultrafast Transmission Electron Microscope, allowing for \textit{in-situ} optical excitation~\cite{Feist2017, Eggebrecht2017, Moeller2020}. We recorded energy-filtered images using a CEOS CEFID equipped with a TVIPS XF416 at a \SI{10}{\electronvolt} window-width,  an acceleration voltage of \SI{200}{\kilo\electronvolt}  and a \SI{6}{\milli\meter} defocus. In these experiments a laser source with a wavelength of \SI{750}{nm} and a pulse width of around \SI{300}{\femto\second} excites the sample under a near normal angle of incidence with a defined number of laser pulses starting from single shot. At various oop magnetic fields, micrographs were acquired before, after a single and after multiple excitations to observe changes of the samples magnetic texture.  In order to obtain reproducible magnetic textures before laser excitation, we fully saturated the sample with an oop field before approaching a specific field magnitude.

In addition to the somewhat different laser conditions, the LTEM sample on the thin TEM membrane has different thermal couplings and strain compared to the MOKE sample on the rigid Si substrate. This also leads to slightly higher magnetic field ranges, where the stripe phase, B/SK lattice phase, and magnetic saturation are present. Due to these differences in experimental conditions, a quantitative fluence matching between TR-MOKE and LTEM experiments was not attempted. However, the LTEM results, at a fluence about a factor of 2 higher than in TR-MOKE, allow to fully understand the the changes to the magnetic texture by laser excitation and allow to perfectly explain the observations in the TR-MOKE measurements. To obtain the fluences in the LTEM measurements from the laser pulse energy, a $1/e^2$ beam diameter of \SI{60}{\micro \meter} was used, estimated from the maximum region of switched magnetic texture upon laser excitation.

\subsection{Static micromagnetic simulations} We use magnum.np~\cite{Bruckner2023}, a GPU-enhanced micromagnetic simulation software, to perform large-scale micromagnetic simulations. We simulate the Fe/Gd multilayers as a 3D ferromagnet with low saturation magnetization $M_s = \SI{340}{kA/m}$ and low perpendicular magnetic anisotropy $K_u = \SI{40}{kJ/m^3}$. The exchange constant is chosen $A_{\mathrm{ex}} = \SI{6}{pJ/m}$. Our finite-difference micromagnetic solver discretizes the simulation box into $512 \times 512 \times 10$ cuboids, each with a constant volume of $10 \times 10 \times 11.2 \mathrm{nm}^3$. Note that all simulations are performed without any thermal fluctuations. Basically, the system is simulated at $T=\SI{0}{K}$, but with effective magnetic parameters that were measured at room temperature. Similar modeling was used in experimental studies on Fe/Gd-based multilayers~\cite{Montoya2017a,Heigl2021}. To obtain the $M$-$H$ hysteresis loops shown in Fig.~\ref{fig:LTEM}, we start from a cellwise random magnetization state. We relax the structure in zero field by solving the Landau-Lifshitz-Gilbert (LLG) equation numerically~\cite{Abert2019} at high damping $\alpha = 1.0$, where we include the demagnetization, exchange, anisotropy and Zeeman energies in the calculation of the effective field. We obtain the stripe domain structure as our new initial state. Once the magnetic system is relaxed with respect to its energy, we apply an oop field along the $z$-direction. We increase the magnetic field by $\SI{2}{mT}$ after each relaxation step in which we solve the LLG for $\SI{10}{ns}$. The complete hysteresis is then simulated by raising and lowering the field between $\pm \SI{250}{mT}$. Magnetization states at each field are saved to be used as input for dynamic simulations to simulate the resonance of the spin objects.

\subsection{Dynamic micromagnetic simulations} With the magnetization states obtained from the static simulations, we now simulate the intrinsic response of the spin textures.  Commonly, ultrafast laser demagnetization can be described micromagnetically by including a stochastic temperature-dependent term to the LLG, where the temperature in the magnetic system is calculated from the excitation energy via the two-temperature model~\cite{Atxitia2007}. However, in our study we are not interested in the demagnetization process itself, but rather in the excitation of the magnetic system and how this relaxes into its equilibrium state. Thus, we employ a similar approach as applied for simulating ferromagnetic resonance (FMR)~\cite{Baker2017}. That is, we assume that the laser pulse heats up the system and alters the temperature-dependent magnetic parameters. Excitation of the system occurs in the system trying to return to its original energetic minimum. Therefore, we assume that during pumping $M_s$ is reduced with an excitation amplitude between $5\%$ and $25\%$. $K_u$ and $A_{\mathrm{ex}}$ scale with $M_s^2$~\cite{Moreno2016}. We start from the magnetization states from the zero-temperature hysteresis at a defined field. We solve the LLG for $\SI{1}{ns}$ at moderate damping $\alpha = 0.1$ to get the magnetization near its new energetic minimum for the new set of magnetic parameters. We then abruptly set the initial magnetic parameters back and reduce the damping to $\alpha = 0.02$. The simulation time is \SI{10}{ns}. The magnetization starts to oscillate back into the original or new ground state, based on the strength of the excitation. Total resonances can then be captured by first recording the averaged magnetization components for every $\Delta t = \SI{1}{ps}$ and performing an FFT. To probe the influence of a stroboscopic measurement technique, we repeat the excitation up to ten times and record the magnetization state after each excitation, where changes in the magnetization states can be observed for high excitation amplitudes. The resonance frequencies of the stripe domain and the skyrmion are separated by averaging the magnetization in a smaller simulation box. For the skyrmion phase we average $m_z$ for a single skyrmion and record it for each time step $\Delta t = \SI{10}{ps}$. For the stripe domain we first find an appropriately small section where only one stripe domain is stable and has almost parallel domain walls. We then sampled the magnetization along a line that is \SI{500}{nm} long. The magnetization is then averaged for each component at the same time step as used for the skyrmions.

\subsection{Data availability}
The data that support the findings of this study are available from the authors upon reasonable request.

\subsection{Code availability}
Micromagnetic simulations were performed with the open source micromagnetic simulator magnum.np~\cite{Bruckner2023}. Simulations scripts to reproduce the results can be requested from corresponding authors.

\subsection{Acknowledgements} S.K.\,and C.A.\,gratefully acknowledge the Austrian Science Fund (FWF) for support through Grant No.\,P34671 (Vladimir). S.K. and D.Su. acknowledge the Austrian Science Fund (FWF) for support through Grant No.\,I 6267 (CHIRALSPIN). T.S. and M.A. gratefully acknowledge funding from Deutsche Forschungsgemeinschaft (DFG, German Research Foundation) grant no.\,507821284. D.St., S.M.\,and T.T.\,acknowledge funding by the DFG, grant no.\,217133147/SFB 1073, project A02. The computational results presented have been achieved in part using the Vienna Scientific Cluster (VSC).

\subsection{Author contributions}
D.St., M.A., and S.M.\,conceived the project. T.T.\,performed the time-resolved Kerr effect studies and analyzed the time-resolved magnetism data with supervision from D.St. Sample growth, static SQUID-VSM and LTEM analysis were performed by T.S. LTEM with femtosecond laser excitation was conceived by C.R., measured by M.M. and T.T., and evaluated by T.T. S.K., F.B., and C.A.\,wrote and improved the micromagnetic code. S.K.\,performed the micromagnetic simulations under the supervision of C.A.\,and D.Su. The simulations were evaluated by S.K. and\,T.T. with support from D.St. T.T., D.St.\,and S.M.\,wrote the paper with input from all authors.

\subsection{Corresponding authors}

Correspondence to Daniel Steil or Stefan Mathias

\subsection{Competing interests}
The authors declare no competing interests.

\bibliographystyle{naturemag}
\bibliography{references}

\begin{thebibliography}{10}
\expandafter\ifx\csname url\endcsname\relax
  \def\url#1{\texttt{#1}}\fi
\expandafter\ifx\csname urlprefix\endcsname\relax\def\urlprefix{URL }\fi
\providecommand{\bibinfo}[2]{#2}
\providecommand{\eprint}[2][]{\url{#2}}

\bibitem{Finocchio2016}
\bibinfo{author}{Finocchio, G.}, \bibinfo{author}{B{\"u}ttner, F.},
  \bibinfo{author}{Tomasello, R.}, \bibinfo{author}{Carpentieri, M.} \&
  \bibinfo{author}{Kl{\"a}ui, M.}
\newblock \bibinfo{title}{Magnetic skyrmions: from fundamental to
  applications}.
\newblock \emph{\bibinfo{journal}{Journal of Physics D: Applied Physics}}
  \textbf{\bibinfo{volume}{49}}, \bibinfo{pages}{423001}
  (\bibinfo{year}{2016}).

\bibitem{Fert2017}
\bibinfo{author}{Fert, A.}, \bibinfo{author}{Reyren, N.} \&
  \bibinfo{author}{Cros, V.}
\newblock \bibinfo{title}{Magnetic skyrmions: advances in physics and potential
  applications}.
\newblock \emph{\bibinfo{journal}{Nature Reviews Materials}}
  \textbf{\bibinfo{volume}{2}}, \bibinfo{pages}{1--15} (\bibinfo{year}{2017}).

\bibitem{Prychynenko2018}
\bibinfo{author}{Prychynenko, D.} \emph{et~al.}
\newblock \bibinfo{title}{Magnetic skyrmion as a nonlinear resistive element: a
  potential building block for reservoir computing}.
\newblock \emph{\bibinfo{journal}{Physical Review Applied}}
  \textbf{\bibinfo{volume}{9}}, \bibinfo{pages}{014034} (\bibinfo{year}{2018}).

\bibitem{Song2020}
\bibinfo{author}{Song, K.~M.} \emph{et~al.}
\newblock \bibinfo{title}{Skyrmion-based artificial synapses for neuromorphic
  computing}.
\newblock \emph{\bibinfo{journal}{Nature Electronics}}
  \textbf{\bibinfo{volume}{3}}, \bibinfo{pages}{148--155}
  (\bibinfo{year}{2020}).

\bibitem{Zhang2020}
\bibinfo{author}{Zhang, X.} \emph{et~al.}
\newblock \bibinfo{title}{Skyrmion-electronics: writing, deleting, reading and
  processing magnetic skyrmions toward spintronic applications}.
\newblock \emph{\bibinfo{journal}{Journal of Physics: Condensed Matter}}
  \textbf{\bibinfo{volume}{32}}, \bibinfo{pages}{143001}
  (\bibinfo{year}{2020}).

\bibitem{Chumak2022}
\bibinfo{author}{Chumak, A.~V.} \emph{et~al.}
\newblock \bibinfo{title}{Advances in magnetics roadmap on spin-wave
  computing}.
\newblock \emph{\bibinfo{journal}{IEEE Transactions on Magnetics}}
  \textbf{\bibinfo{volume}{58}}, \bibinfo{pages}{1--72} (\bibinfo{year}{2022}).

\bibitem{Bogdanov1994}
\bibinfo{author}{Bogdanov, A.} \& \bibinfo{author}{Hubert, A.}
\newblock \bibinfo{title}{Thermodynamically stable magnetic vortex states in
  magnetic crystals}.
\newblock \emph{\bibinfo{journal}{Journal of magnetism and magnetic materials}}
  \textbf{\bibinfo{volume}{138}}, \bibinfo{pages}{255--269}
  (\bibinfo{year}{1994}).

\bibitem{Roessler2006}
\bibinfo{author}{Roessler, U.~K.}, \bibinfo{author}{Bogdanov, A.} \&
  \bibinfo{author}{Pfleiderer, C.}
\newblock \bibinfo{title}{Spontaneous skyrmion ground states in magnetic
  metals}.
\newblock \emph{\bibinfo{journal}{Nature}} \textbf{\bibinfo{volume}{442}},
  \bibinfo{pages}{797--801} (\bibinfo{year}{2006}).

\bibitem{Yu2010}
\bibinfo{author}{Yu, X.} \emph{et~al.}
\newblock \bibinfo{title}{Real-space observation of a two-dimensional skyrmion
  crystal}.
\newblock \emph{\bibinfo{journal}{Nature}} \textbf{\bibinfo{volume}{465}},
  \bibinfo{pages}{901--904} (\bibinfo{year}{2010}).

\bibitem{Lenk2011}
\bibinfo{author}{Lenk, B.}, \bibinfo{author}{Ulrichs, H.},
  \bibinfo{author}{Garbs, F.} \& \bibinfo{author}{Münzenberg, M.}
\newblock \bibinfo{title}{The building blocks of magnonics}.
\newblock \emph{\bibinfo{journal}{Physics Reports}}
  \textbf{\bibinfo{volume}{507}}, \bibinfo{pages}{107--136}
  (\bibinfo{year}{2011}).

\bibitem{Yu2021}
\bibinfo{author}{Yu, H.}, \bibinfo{author}{Xiao, J.} \&
  \bibinfo{author}{Schultheiss, H.}
\newblock \bibinfo{title}{Magnetic texture based magnonics}.
\newblock \emph{\bibinfo{journal}{Physics Reports}}
  \textbf{\bibinfo{volume}{905}}, \bibinfo{pages}{1--59}
  (\bibinfo{year}{2021}).

\bibitem{Petti2022}
\bibinfo{author}{Petti, D.}, \bibinfo{author}{Tacchi, S.} \&
  \bibinfo{author}{Albisetti, E.}
\newblock \bibinfo{title}{Review on magnonics with engineered spin textures}.
\newblock \emph{\bibinfo{journal}{Journal of Physics D: Applied Physics}}
  \textbf{\bibinfo{volume}{55}}, \bibinfo{pages}{293003}
  (\bibinfo{year}{2022}).

\bibitem{Ogawa2015}
\bibinfo{author}{Ogawa, N.}, \bibinfo{author}{Seki, S.} \&
  \bibinfo{author}{Tokura, Y.}
\newblock \bibinfo{title}{Ultrafast optical excitation of magnetic skyrmions}.
\newblock \emph{\bibinfo{journal}{Scientific Reports}}
  \textbf{\bibinfo{volume}{5}}, \bibinfo{pages}{9552} (\bibinfo{year}{2015}).

\bibitem{Padmanabhan2019}
\bibinfo{author}{Padmanabhan, P.} \emph{et~al.}
\newblock \bibinfo{title}{Optically driven collective spin excitations and
  magnetization dynamics in the {N}\'eel-type skyrmion host
  {G}a{V}$_{4}${S}$_{8}$}.
\newblock \emph{\bibinfo{journal}{Phys. Rev. Lett.}}
  \textbf{\bibinfo{volume}{122}}, \bibinfo{pages}{107203}
  (\bibinfo{year}{2019}).

\bibitem{Sekiguchi2022}
\bibinfo{author}{Sekiguchi, F.} \emph{et~al.}
\newblock \bibinfo{title}{Slowdown of photoexcited spin dynamics in the
  non-collinear spin-ordered phases in skyrmion host {Ga}{V}$_4${S}$_8$}.
\newblock \emph{\bibinfo{journal}{Nature Communications}}
  \textbf{\bibinfo{volume}{13}}, \bibinfo{pages}{3212} (\bibinfo{year}{2022}).

\bibitem{Kalin2022}
\bibinfo{author}{Kalin, J.} \emph{et~al.}
\newblock \bibinfo{title}{Optically excited spin dynamics of thermally
  metastable skyrmions in {F}e$_{0.75}${C}o$_{0.25}$}.
\newblock \emph{\bibinfo{journal}{Phys. Rev. B}}
  \textbf{\bibinfo{volume}{106}}, \bibinfo{pages}{054430}
  (\bibinfo{year}{2022}).

\bibitem{Eggebrecht2017}
\bibinfo{author}{Eggebrecht, T.} \emph{et~al.}
\newblock \bibinfo{title}{Light-induced metastable magnetic texture uncovered
  by in situ {L}orentz microscopy}.
\newblock \emph{\bibinfo{journal}{Phys. Rev. Lett.}}
  \textbf{\bibinfo{volume}{118}}, \bibinfo{pages}{097203}
  (\bibinfo{year}{2017}).

\bibitem{Je2018}
\bibinfo{author}{Je, S.-G.} \emph{et~al.}
\newblock \bibinfo{title}{Creation of magnetic skyrmion bubble lattices by
  ultrafast laser in ultrathin films}.
\newblock \emph{\bibinfo{journal}{Nano Letters}} \textbf{\bibinfo{volume}{18}},
  \bibinfo{pages}{7362--7371} (\bibinfo{year}{2018}).

\bibitem{Buettner2021}
\bibinfo{author}{B{\"u}ttner, F.} \emph{et~al.}
\newblock \bibinfo{title}{{Observation of fluctuation-mediated picosecond
  nucleation of a topological phase}}.
\newblock \emph{\bibinfo{journal}{{Nature Materials}}}
  \textbf{\bibinfo{volume}{20}}, \bibinfo{pages}{30--37}
  (\bibinfo{year}{2021}).

\bibitem{Khela2023}
\bibinfo{author}{Khela, M.} \emph{et~al.}
\newblock \bibinfo{title}{Laser-induced topological spin switching in a 2{D}
  van der waals magnet}.
\newblock \emph{\bibinfo{journal}{Nature Communications}}
  \textbf{\bibinfo{volume}{14}}, \bibinfo{pages}{1378} (\bibinfo{year}{2023}).

\bibitem{Montoya2017a}
\bibinfo{author}{Montoya, S.~A.} \emph{et~al.}
\newblock \bibinfo{title}{Tailoring magnetic energies to form dipole skyrmions
  and skyrmion lattices}.
\newblock \emph{\bibinfo{journal}{Phys. Rev. B}} \textbf{\bibinfo{volume}{95}},
  \bibinfo{pages}{024415} (\bibinfo{year}{2017}).

\bibitem{Heigl2021}
\bibinfo{author}{Heigl, M.} \emph{et~al.}
\newblock \bibinfo{title}{Dipolar-stabilized first and second-order
  antiskyrmions in ferrimagnetic multilayers}.
\newblock \emph{\bibinfo{journal}{Nat Commun}} \textbf{\bibinfo{volume}{12}},
  \bibinfo{pages}{2611} (\bibinfo{year}{2021}).

\bibitem{Beaurepaire1996}
\bibinfo{author}{Beaurepaire, E.}, \bibinfo{author}{Merle, J.-C.},
  \bibinfo{author}{Daunois, A.} \& \bibinfo{author}{Bigot, J.-Y.}
\newblock \bibinfo{title}{Ultrafast spin dynamics in ferromagnetic nickel}.
\newblock \emph{\bibinfo{journal}{Phys. Rev. Lett.}}
  \textbf{\bibinfo{volume}{76}}, \bibinfo{pages}{4250--4253}
  (\bibinfo{year}{1996}).

\bibitem{Kirilyuk2010}
\bibinfo{author}{Kirilyuk, A.}, \bibinfo{author}{Kimel, A.~V.} \&
  \bibinfo{author}{Rasing, T.}
\newblock \bibinfo{title}{Ultrafast optical manipulation of magnetic order}.
\newblock \emph{\bibinfo{journal}{Rev. Mod. Phys.}}
  \textbf{\bibinfo{volume}{82}}, \bibinfo{pages}{2731--2784}
  (\bibinfo{year}{2010}).

\bibitem{Mochizuki2012}
\bibinfo{author}{Mochizuki, M.}
\newblock \bibinfo{title}{Spin-wave modes and their intense excitation effects
  in skyrmion crystals}.
\newblock \emph{\bibinfo{journal}{Phys. Rev. Lett.}}
  \textbf{\bibinfo{volume}{108}}, \bibinfo{pages}{017601}
  (\bibinfo{year}{2012}).

\bibitem{Onose2012}
\bibinfo{author}{Onose, Y.}, \bibinfo{author}{Okamura, Y.},
  \bibinfo{author}{Seki, S.}, \bibinfo{author}{Ishiwata, S.} \&
  \bibinfo{author}{Tokura, Y.}
\newblock \bibinfo{title}{Observation of magnetic excitations of skyrmion
  crystal in a helimagnetic insulator {C}u$_{2}${O}{S}e{O}$_{3}$}.
\newblock \emph{\bibinfo{journal}{Phys. Rev. Lett.}}
  \textbf{\bibinfo{volume}{109}}, \bibinfo{pages}{037603}
  (\bibinfo{year}{2012}).

\bibitem{Koshibae2014}
\bibinfo{author}{Koshibae, W.} \& \bibinfo{author}{Nagaosa, N.}
\newblock \bibinfo{title}{Creation of skyrmions and antiskyrmions by local
  heating}.
\newblock \emph{\bibinfo{journal}{Nature Communications}}
  \textbf{\bibinfo{volume}{5}}, \bibinfo{pages}{5148} (\bibinfo{year}{2014}).

\bibitem{Gerlinger2021}
\bibinfo{author}{Gerlinger, K.} \emph{et~al.}
\newblock \bibinfo{title}{Application concepts for ultrafast laser-induced
  skyrmion creation and annihilation}.
\newblock \emph{\bibinfo{journal}{Applied Physics Letters}}
  \textbf{\bibinfo{volume}{118}}, \bibinfo{pages}{192403}
  (\bibinfo{year}{2021}).

\bibitem{moller_ultrafast_2021}
\bibinfo{author}{Möller, C.} \emph{et~al.}
\newblock \bibinfo{title}{Ultrafast element-resolved magneto-optics using a
  fiber-laser-driven extreme ultraviolet light source}.
\newblock \emph{\bibinfo{journal}{Review of Scientific Instruments}}
  \textbf{\bibinfo{volume}{92}}, \bibinfo{pages}{065107}
  (\bibinfo{year}{2021}).

\bibitem{Feist2017}
\bibinfo{author}{Feist, A.} \emph{et~al.}
\newblock \bibinfo{title}{Ultrafast transmission electron microscopy using a
  laser-driven field emitter: Femtosecond resolution with a high coherence
  electron beam}.
\newblock \emph{\bibinfo{journal}{Ultramicroscopy}}
  \textbf{\bibinfo{volume}{176}}, \bibinfo{pages}{63--73}
  (\bibinfo{year}{2017}).

\bibitem{Moeller2020}
\bibinfo{author}{Möller, M.}, \bibinfo{author}{Gaida, J.~H.},
  \bibinfo{author}{Schäfer, S.} \& \bibinfo{author}{Ropers, C.}
\newblock \bibinfo{title}{Few-nm tracking of current-driven magnetic vortex
  orbits using ultrafast {L}orentz microscopy}.
\newblock \emph{\bibinfo{journal}{Communications Physics}}
  \textbf{\bibinfo{volume}{3}}, \bibinfo{pages}{36} (\bibinfo{year}{2020}).

\bibitem{Bruckner2023}
\bibinfo{author}{Bruckner, F.}, \bibinfo{author}{Koraltan, S.},
  \bibinfo{author}{Abert, C.} \& \bibinfo{author}{Suess, D.}
\newblock \bibinfo{title}{magnum.np: a {P}y{T}orch based {GPU} enhanced finite
  difference micromagnetic simulation framework for high level development and
  inverse design}.
\newblock \emph{\bibinfo{journal}{Scientific Reports}}
  \textbf{\bibinfo{volume}{13}}, \bibinfo{pages}{12054} (\bibinfo{year}{2023}).

\bibitem{Abert2019}
\bibinfo{author}{Abert, C.}
\newblock \bibinfo{title}{Micromagnetics and spintronics: models and numerical
  methods}.
\newblock \emph{\bibinfo{journal}{The European Physical Journal B}}
  \textbf{\bibinfo{volume}{92}}, \bibinfo{pages}{120} (\bibinfo{year}{2019}).

\bibitem{Atxitia2007}
\bibinfo{author}{Atxitia, U.} \emph{et~al.}
\newblock \bibinfo{title}{{Micromagnetic modeling of laser-induced
  magnetization dynamics using the {L}andau-{L}ifshitz-{B}loch equation}}.
\newblock \emph{\bibinfo{journal}{Applied Physics Letters}}
  \textbf{\bibinfo{volume}{91}}, \bibinfo{pages}{232507}
  (\bibinfo{year}{2007}).

\bibitem{Baker2017}
\bibinfo{author}{Baker, A.} \emph{et~al.}
\newblock \bibinfo{title}{Proposal of a micromagnetic standard problem for
  ferromagnetic resonance simulations}.
\newblock \emph{\bibinfo{journal}{Journal of Magnetism and Magnetic Materials}}
  \textbf{\bibinfo{volume}{421}}, \bibinfo{pages}{428--439}
  (\bibinfo{year}{2017}).

\bibitem{Moreno2016}
\bibinfo{author}{Moreno, R.} \emph{et~al.}
\newblock \bibinfo{title}{Temperature-dependent exchange stiffness and domain
  wall width in {C}o}.
\newblock \emph{\bibinfo{journal}{Phys. Rev. B}} \textbf{\bibinfo{volume}{94}},
  \bibinfo{pages}{104433} (\bibinfo{year}{2016}).

\bibitem{Bayer2005}
\bibinfo{author}{Bayer, C.}, \bibinfo{author}{Schultheiss, H.},
  \bibinfo{author}{Hillebrands, B.} \& \bibinfo{author}{Stamps, R.~L.}
\newblock \bibinfo{title}{Phase shift of spin waves traveling through a
  180$^\circ$ bloch-domain wall}.
\newblock \emph{\bibinfo{journal}{IEEE Transactions on Magnetics}}
  \textbf{\bibinfo{volume}{41}}, \bibinfo{pages}{3094--3096}
  (\bibinfo{year}{2005}).

\end{thebibliography}

\section{Supplementary}
\setcounter{figure}{0}
\renewcommand{\figurename}{Fig.}
\renewcommand{\thefigure}{S\arabic{figure}}
\subsection{Weak perturbative regime: Extended discussion of the spin dynamics of stripe domains, bubbles, and skyrmions}\label{supp_pert}
This section provides an extended comparison of the experimental results and micromagnetic simulations in the weak perturbation limit. We also provide further information on the different observed breathing modes by comparison of experiment and simulation.

%We further want to refer the reader to movies of the breathing of stripes and skyrmions from micromagnetic simulations, which are provided as separate files.

\subsubsection{Field-dependent differences in coherent spin dynamics between experiment and simulation}
For magnetic fields $<\SI{150}{\milli \tesla}$ the results obtained from modelling and experiment in Fig.~\ref{fig:oscTheoExp} are in very good agreement. The mode $m_1$ with frequency $f_1\approx\SI{2.3}{\giga \hertz}$ is observed in both cases. However, the higher-frequency mode $m_2$ as well as the opposite phase shift observed in experiment for up- and downsweep of the magnetic field are not reproduced by the simulation. We attribute these differences to sample properties not accounted for in the simulation, e.g., sample inhomogeneities and defects as well as the fact that the simulation is a strongly discretized situation.

Furthermore, in the magnetic field range $\SI{150}{\milli \tesla}<\mu_0 H<\SI{200}{\milli \tesla}$ a sharp transition exists between the two modes $m_1$ and $m_{bsk}$ in the experiment, while in the simulation this transition appears to be more continuous. Additionally, a transition area for mode $m_1$ is also present for the downsweep of the magnetic field in the simulation. We explain this difference to the experiment with the high amount of cylindrical-like spin objects in the simulated ground state of the mixed phase for both magnetic field up- and downsweep, see micromagnetic simulation data in Fig.~\ref{fig:LTEM}b, image III. In contrast, according to the LTEM data, the ground state in the experiment is mainly governed by stripe domains. Due to the strong spatial discretization within the simulation, as well as the absence of defects likely acting as nucleation sites for domain walls in the actual sample, the evolution of spin texture, especially starting from a formerly saturated state, can be expected to deviate from the experiment.

\subsubsection{Breathing mode of a single skyrmion}
\begin{figure}[h!]
     \centering
     \includegraphics[width=\columnwidth]{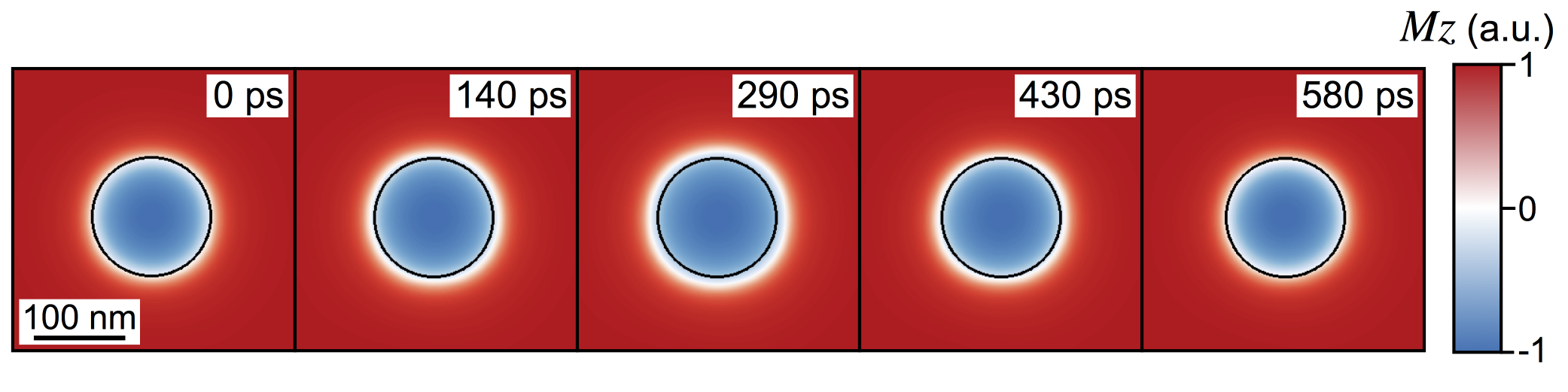}
     \caption{\textbf{Breathing mode of a single skyrmion.} One period of the breathing mode of a single skyrmion obtained from micromagnetic simulations. The black circle indicates the initial skyrmion size as a guide to the eye.}
     \label{fig:Breathing}
\end{figure}

A snapshot of the spin texture is collected for each time step of the micromagnetic simulation. We select one skyrmion ($\mu_0 H=\SI{200}{\milli \tesla}$, upsweep) to demonstrate the time evolution of the skyrmion breathing mode for a single period in Fig.~\ref{fig:Breathing}.

\subsubsection{Comparison of stripes, bubble/skyrmion breathing modes in simulation and experiment}
\begin{figure}[h!]
     \centering
     \includegraphics[width=0.7\columnwidth]{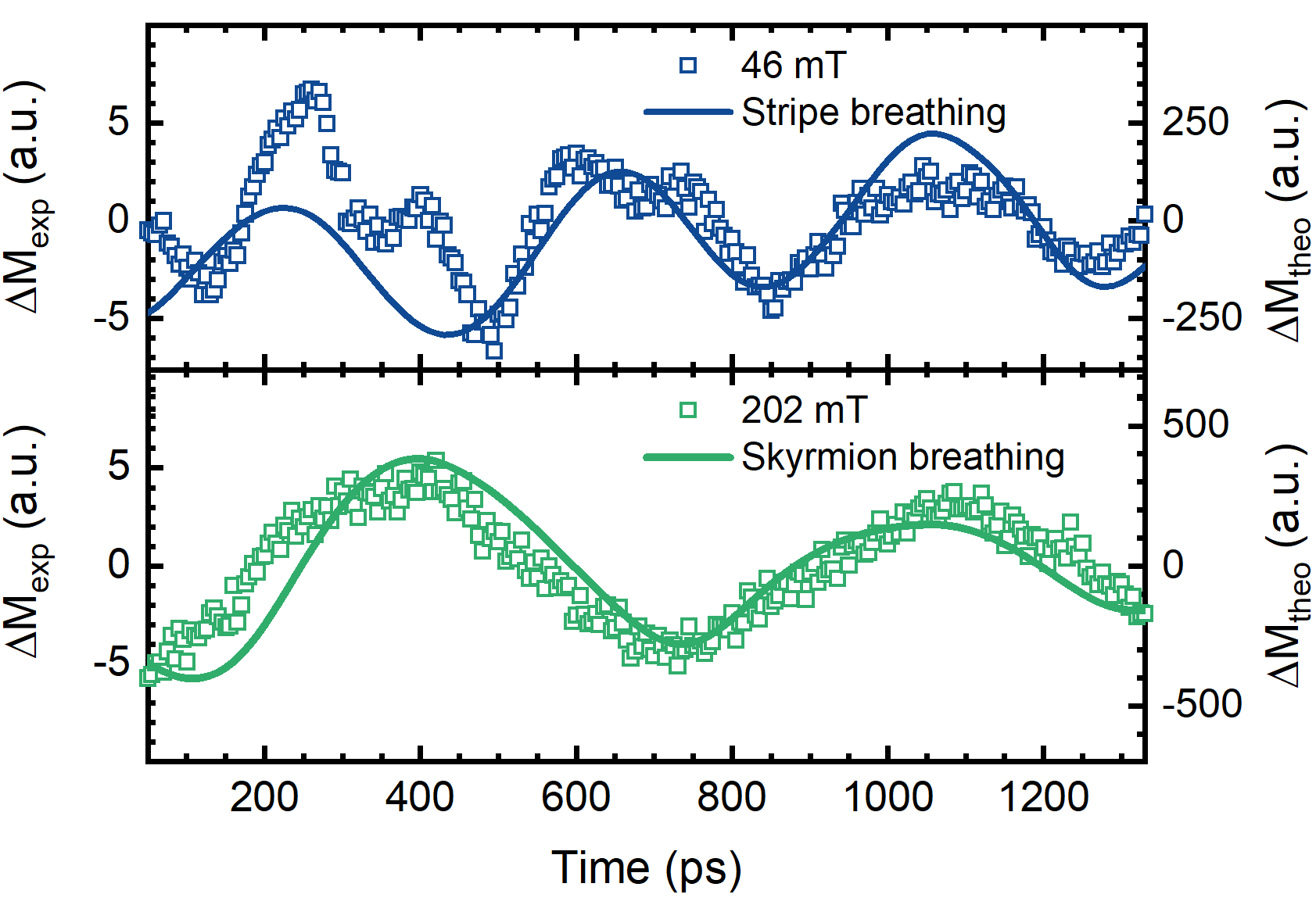}
     \caption{\textbf{Breathing modes of stripes and B/SK.} Low-fluence coherent magnetization dynamics within the stripe and B/SK phase that can be linked to the collective breathing modes of stripes and skyrmions, respectively. Data points are experimental data corresponding to the collective response of the $M_z$-component of the spin system in the probed sample area, solid lines depict the $M_z$-component of the magnetization from the micromagnetic simulations for an area of \SI{300}{}x\SI{300}{\nano \meter}$^2$ containing a single skyrmion (bottom) or \SI{300}{}x\SI{500}{\nano \meter}$^2$ for part of a stripe domain (top).}
     \label{fig:StSk}
\end{figure}

The width of a stripe domain and the size of B/SKs change periodically as a response to the excitation. This breathing mode causes a periodic oscillation of the magnetization component along the oop direction, which can be observed both in our TR-MOKE experiment and in micromagnetic simulations as depicted in Fig.~\ref{fig:StSk}. Note that the experimental data corresponds to the integrated collective spin response over the probed sample area, whereas the simulation (solid lines) shows the integrated $M_z$-component of the magnetization in a small area containing a single skyrmion or a single stripe domain. Especially at early times, deviations are to be expected, because the excitation in experiment and the micromagnetic simulations are different (see Methods section). Furthermore, in the stripe breathing the higher frequency mode $m_2$ is present in the experiment, but not in the simulation. Overall, considering these differences, experiment and theory match closely.

\subsubsection{Phase and frequency shifts in oscillatory spin dynamics}
In general, the spin wave dynamics observed in Fig.~\ref{fig:oscTheoExp} and Fig.~\ref{fig:oscpart200mW} exhibit a field-dependent phase and/or frequency shift both in the stripe domain and B/SK lattice phases. The expected direction of these shifts is currently not well understood by the authors, but is believed to partially result from the density of domain walls that will influence the spin wave phase~\cite{Bayer2005}. We discuss below two particular examples from Fig.~\ref{fig:oscTheoExp} and outline our current understanding of the observed effect.

\paragraph{B/SK lattice phase shift---weak perturbation limit}

\begin{figure}[htb]
     \centering
     \includegraphics[width=\columnwidth]{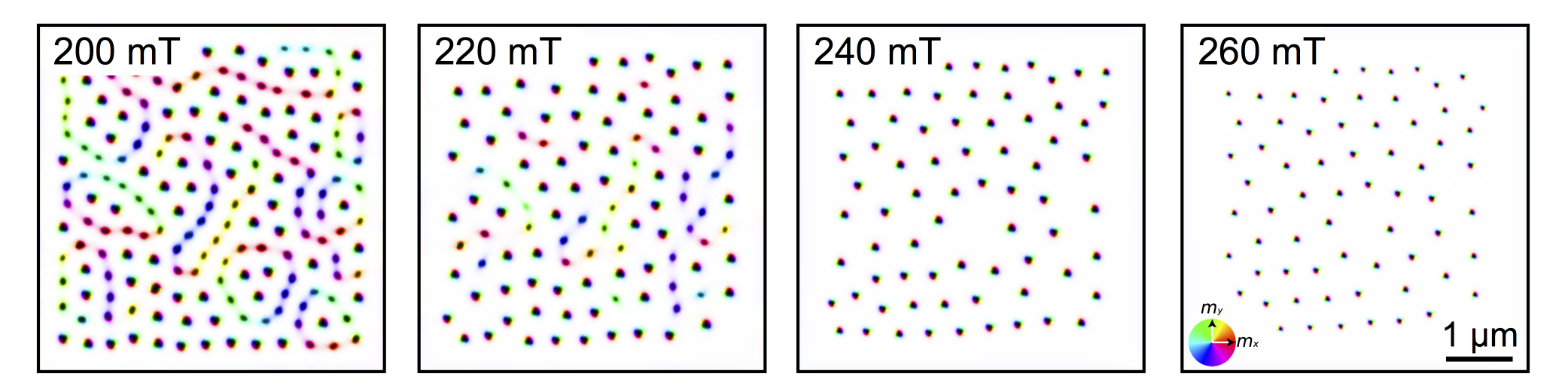}
     \caption{\textbf{Micromagnetic simulation of field-dependent B/SK evolution.} Evolution of the size of cylindrical spin objects with magnetic field in the micromagnetic simulation. The direction of $M$ is given by the color code.}
     \label{fig:SkyrmionSize}
\end{figure}

The frequency of $m_{bsk}$ does not visibly depend on the oop magnetic field in experiment and simulation. However, we experimentally observe a shift of the oscillation onset by $\Delta t \approx \SI{100}{\pico \second}$ with increasing magnetic field and a phase shift is also visible in the simulation. Here, this phase shift interestingly appears together with the magnetic field-dependent vanishing of bubbles and shrinking of skyrmions.
As depicted in Fig.~\ref{fig:SkyrmionSize}, the micromagnetic simulation predicts a drop in both size and number of cylindrical spin objects with respect to an increasing applied magnetic field. As a result, the amount of Bloch-like domain walls is reduced. However, the phase of a spin wave is modified when transmitting through a Bloch-like domain wall~\cite{Bayer2005}. Since there are fewer domain walls at higher magnetic fields, the phase shift must change, which we attribute to the observed phase shift of the $m_{bsk}$ mode.

\begin{figure}[htb]
     \centering
     \includegraphics[width=\columnwidth]{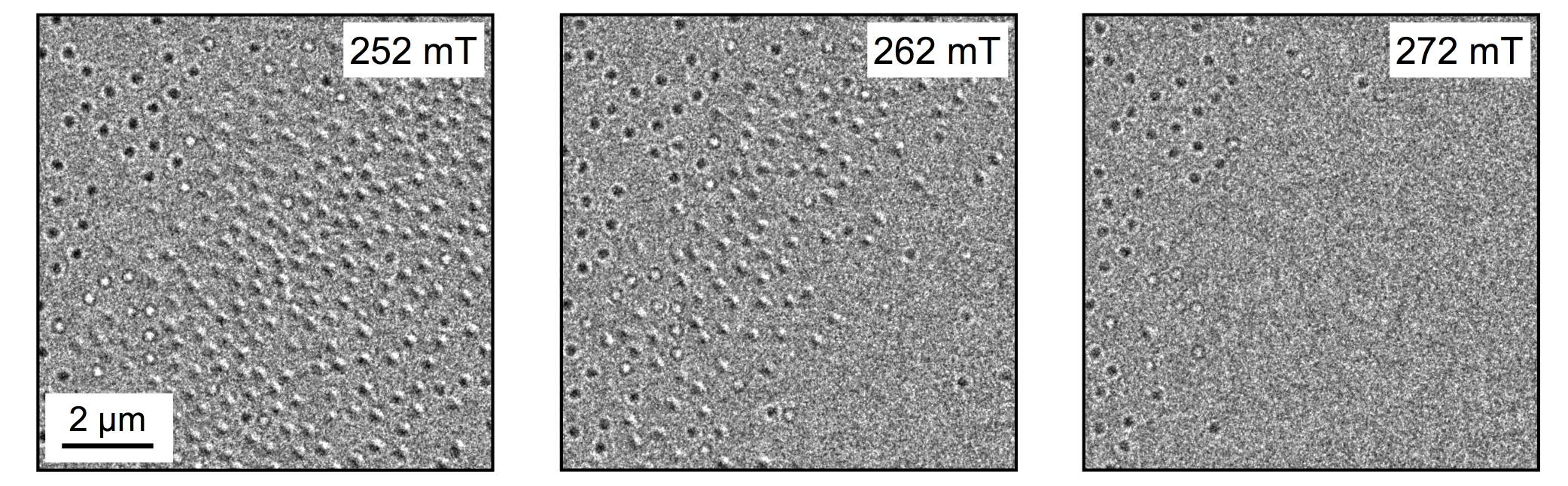}
     \caption{\textbf{LTEM measurement of field-dependent B/SK evolution.} Experimentally measured spin texture in dependence of the magnetic oop field for upsweep of the magnetic field. Bubbles are observed to be less stable than skyrmions.}
     \label{fig:BubblesSkyrmions}
\end{figure}

Experimentally, we find that skyrmions are more stable with respect to an increasing magnetic field than bubbles. As depicted in Fig.~\ref{fig:BubblesSkyrmions}, the bubbles continuously vanish with increasing field while the skyrmions remain almost unaffected. As a result, within a specific magnetic field range, only skyrmions are present. Again, the density of domain walls is reduced, supporting the above argument.

\paragraph{Coherent magnetization dynamics within the mixed phase}

\begin{figure}[htb]
     \centering
     \includegraphics[width=0.7\columnwidth]{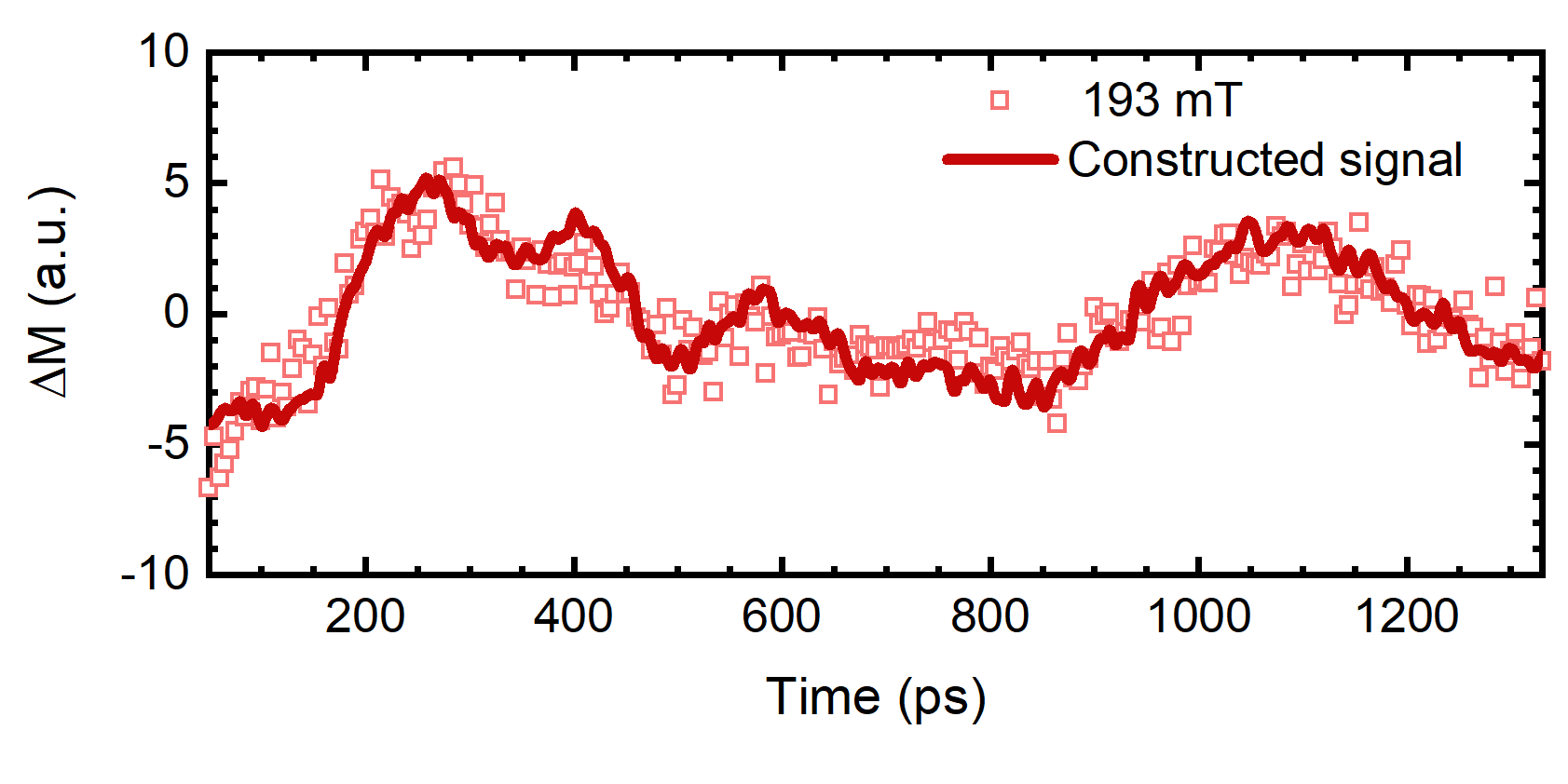}
     \caption{\textbf{Coherent magnetization dynamics in the mixed phase state.} Low-fluence coherent magnetization dynamics within the mixed phase with experimental data as red squares. The dynamics are modelled (solid line) by a superposition of the coherent dynamics within stripe domain and B/SK lattice phase as depicted in Fig.~\ref{fig:StSk}}.
     \label{fig:Superposition}
\end{figure}

Considering the upsweep of the magnetic field in the experiment, we observe a non-monotonous behavior of the observed oscillation shift within the stripe domain phase, e.g., the oscillations shift to the right for magnetic fields $\mu_0 H>\SI{150}{\milli \tesla}$ (see Fig.~\ref{fig:oscTheoExp}a). We attribute this behavior to the coexistence of stripes and B/SKs within this field range. As a result, the coherent dynamics can be modelled as a superposition of the stripe domain and B/SK breathing mode. In Fig.~\ref{fig:Superposition} we exemplarily constructed the coherent dynamics for $\mu_0 H=\SI{193}{\milli \tesla}$ by using the breathing modes shown in Fig.~\ref{fig:StSk} with a B/SK to stripe domain ratio of \SI{1.4}{}. For the strong excitation regime given in Fig.~\ref{fig:oscpart200mW}c)~and~d), this effect manifests in a resolvable frequency shift in the mixed phase region between $\SI{110}{}<\mu_0 H<\SI{150}{\milli \tesla}$.

\subsection{Strong perturbative regime: Extended discussion of the laser-induced transformation of spin textures}

\subsubsection{Magnetization dynamics: Strong optical excitation}
\begin{figure}[htb]
     \centering
     \includegraphics[width=0.7\columnwidth]{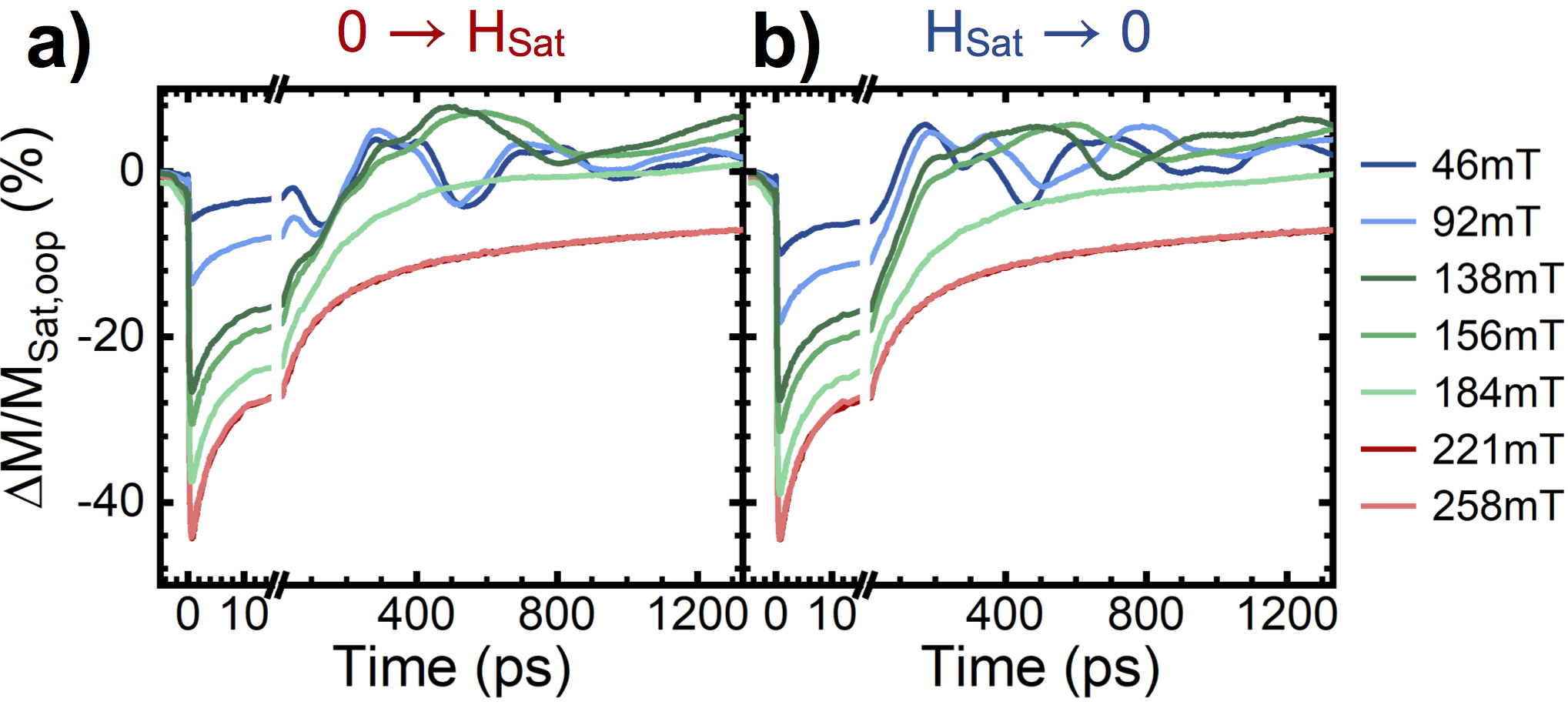}
     \caption{\textbf{Field-dependent TR-MOKE for strong optical excitation.} Out-of-plane magnetization dynamics in dependence of the applied oop magnetic field for upsweep \textbf{a)} and downsweep \textbf{b)} of the magnetic field in the high fluence regime. The colors of the curves represent the attribution to different static spin textures: Stripe domains (blue), bubble/skyrmion lattice (green), and magnetic saturation (red).}
     \label{fig:overview200mW}
\end{figure}

Figure~\ref{fig:overview200mW} depicts the magnetization dynamics for a laser excitation of $F=\SI{3}{\milli \joule \per \centi \meter \squared}$. In comparison to the weak excitation shown in Fig.~\ref{fig:overview20mW}, we observe a much stronger incoherent demagnetization of about 45\%, which is indicative for the strong optically-induced perturbation of the sample.

\subsubsection{Skyrmion nucleation from an isolated chiral domain wall}
\begin{figure}[h!]
     \centering
     \includegraphics[width=0.8\columnwidth]{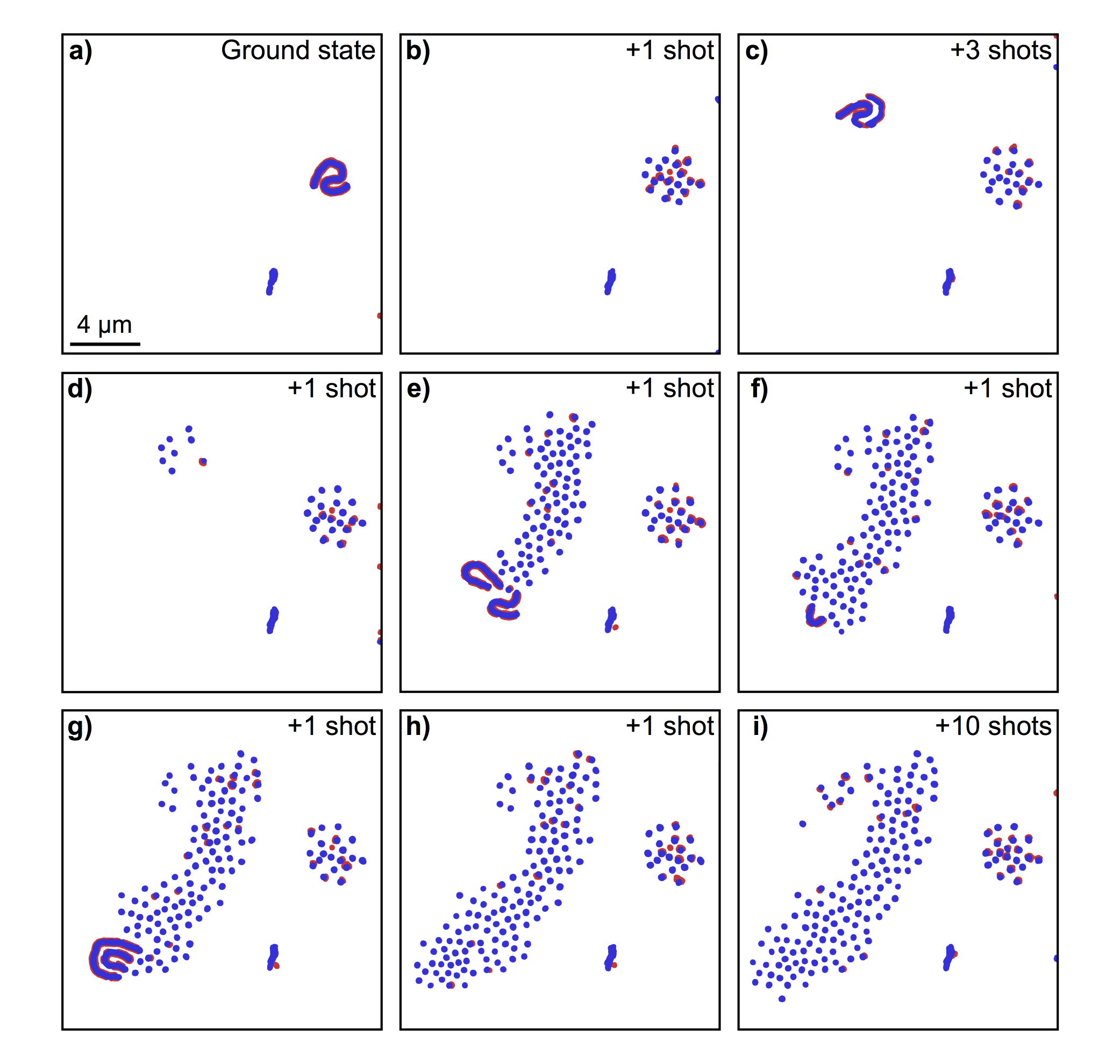}
     \caption{\textbf{Skyrmion nucleation from an isolated chiral domain wall.} From \textbf{a)} to \textbf{i)}: Evolution of spin texture in dependence of the number of laser pulses with fluence of \SI{6.4}{\milli \joule \per \centi \meter \squared}. The magnetic field is chosen to be at the transition from saturation to the stripe domain phase during a downsweep of the magnetic field. Note that the LTEM images were significantly post-processed to enhance the visibility of magnetic skyrmions and their chirality.}
     \label{fig:SkfromSt}
\end{figure}

The process of skyrmion nucleation starting from an almost fully saturated state in the downsweep of the magnetic field is depicted in Fig.~\ref{fig:SkfromSt}~a-i. The LTEM images were significantly post-processed to obtain better visibility of the magnetic texture. In more detail, the contours of chiral spin objects associated with black and white LTEM contrast were extracted by running a computer vision algorithm. Small contours that are referred to image noise were disregarded. The remaining contours were then filled and framed with color to increase the visibility of the spin objects. The color indicates the underlying chirality of the spin object, with blue objects and red objects originating from black and white LTEM contrast, respectively.

For the chosen magnetic field of $\mu_0 H=\SI{210}{\milli \tesla}$ in the downsweep of the magnetic field only a single stripe domain is present. This stripe domain transforms into a local skyrmion lattice upon excitation by the first laser pulse. However, multiple shots are required to nucleate a new chiral stripe domain pointing towards a rather stochastical nature of this process. During the following five shots, we observe a subsequent growth of the generated skyrmion phase that remains almost unchanged after ten additional shots. Interestingly, skyrmions nucleate directly after single shot excitation, however, only in a well defined area adjacent to already existing skyrmions. We assume that this area is preferred for chiral stripe domain nucleation upon a slight decrease of the magnetic field, i.e., the local spin texture is not completely saturated. Laser-induced demagnetization then results in a strong perturbation of this pre-stripe state. However, since the chirality of the pre-stripe can be seen as an additional energy barrier, we assume that chirality is conserved upon excitation. Instead, the spins along the stripe domains are affected and cause local inhomogeneity and a partition of the stripe domains in various parts (cf.\,Fig.~\ref{fig:StSkTrafo}). This spin alignment then leads to a preferred forming of a skyrmion phase during the relaxation process. Therefore, the chirality of all generated skyrmions is identical.

\subsubsection{Optically-induced phase transformations: Schematic phase diagram}
\begin{figure}[htb]
     \centering
     \includegraphics[width=\columnwidth]{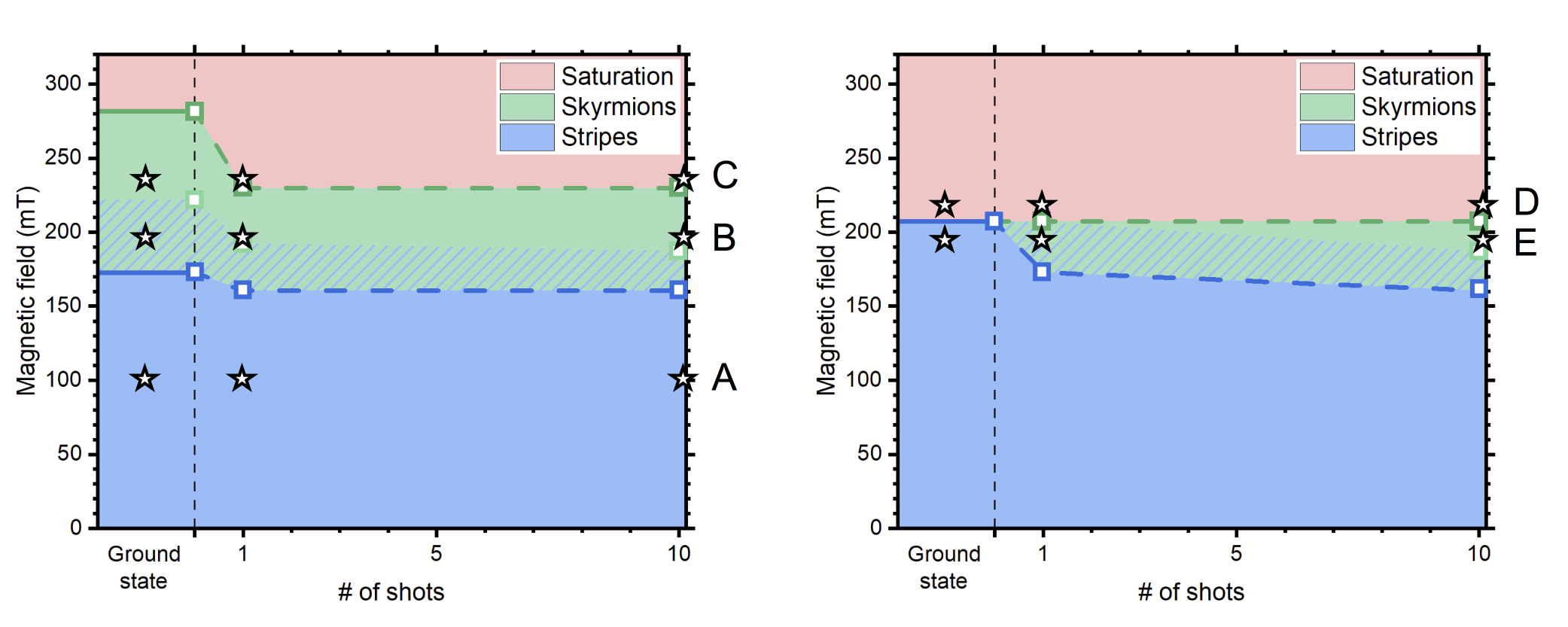}
     \caption{\textbf{Spin texture phase diagram.} Schematic phase diagram of the spin texture in dependence of magnetic field and number of laser pulses for up- a) and downsweep b) of the magnetic field for a fixed fluence of \SI{6.4}{\milli \joule \per \centi \meter \squared}. Stars correspond to the LTEM data depicted in Fig.~\ref{fig:ShotsLTEM}. The magnetic field ranges correspond to the spin textures for the sample on the membrane, which are higher than for the MOKE sample on a Si substrate, see also Fig.~\ref{fig:LTEM}}.
     \label{fig:Phasediagram}
\end{figure}

Figure~\ref{fig:Phasediagram} shows a schematic phase diagram of the magnetic texture depending on both the magnetic field and the number of incident laser pulses. Depending on the magnetic state of the system, we observe different scenarios upon laser excitation including unchanged magnetic texture (A,D), B/SK creation from stripe domains (B,E), and B/SK annihilation (C).

\begin{figure}[h!]
     \centering
     \includegraphics[width=\columnwidth]{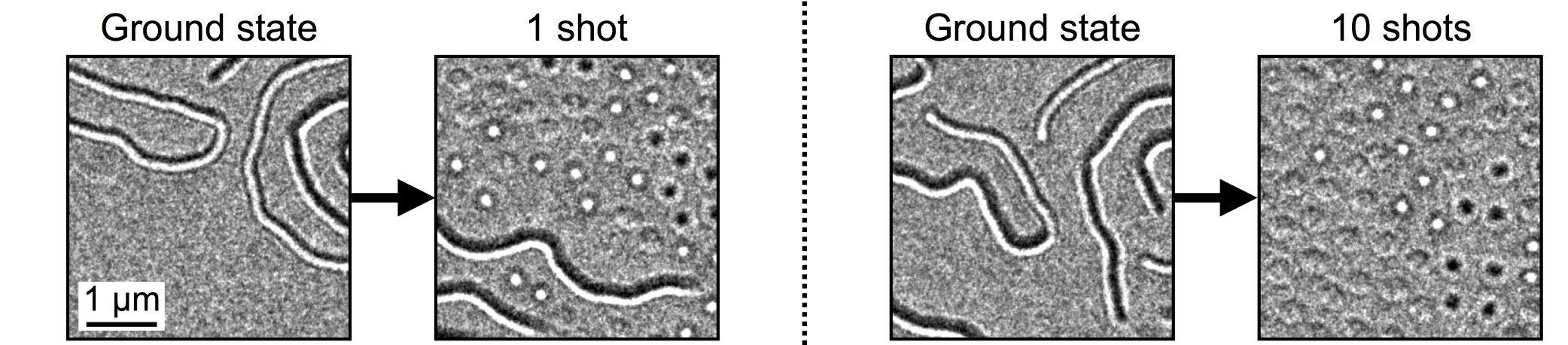}
     \caption{\textbf{Transformation of stripe state into B/SK lattice state via a mixed state.} Left: A stripe state in the downsweep of the magnetic field transforms into a mixed state upon excitation by a single laser pulse with fluence of \SI{6.4}{\milli \joule \per \centi \meter \squared}. Right: Excitation by multiple laser pulses leads to a B/SK lattice state.}
     \label{fig:ShotsLTEMScE}
\end{figure}

In case of scenario E (see also Fig.~\ref{fig:ShotsLTEM}), laser excitation leads to a gradual transformation of a pristine stripe domain phase in case of the downsweep of the magnetic field, as depicted in Fig.~\ref{fig:ShotsLTEMScE}. Upon excitation by a single laser pulse the initial stripes transform into bubbles and skyrmions, respectively. In addition to that, we observe the creation of stripe domains in formerly saturated regions. These generated stripe domains transform in turn into bubbles/skyrmions, finally yielding a pure bubble/skyrmion lattice phase in the laser excited region.

\subsubsection{Skyrmion nucleation in the micromagnetic simulations}
\begin{figure}[htb]
     \centering
     \includegraphics[width=0.8\columnwidth]{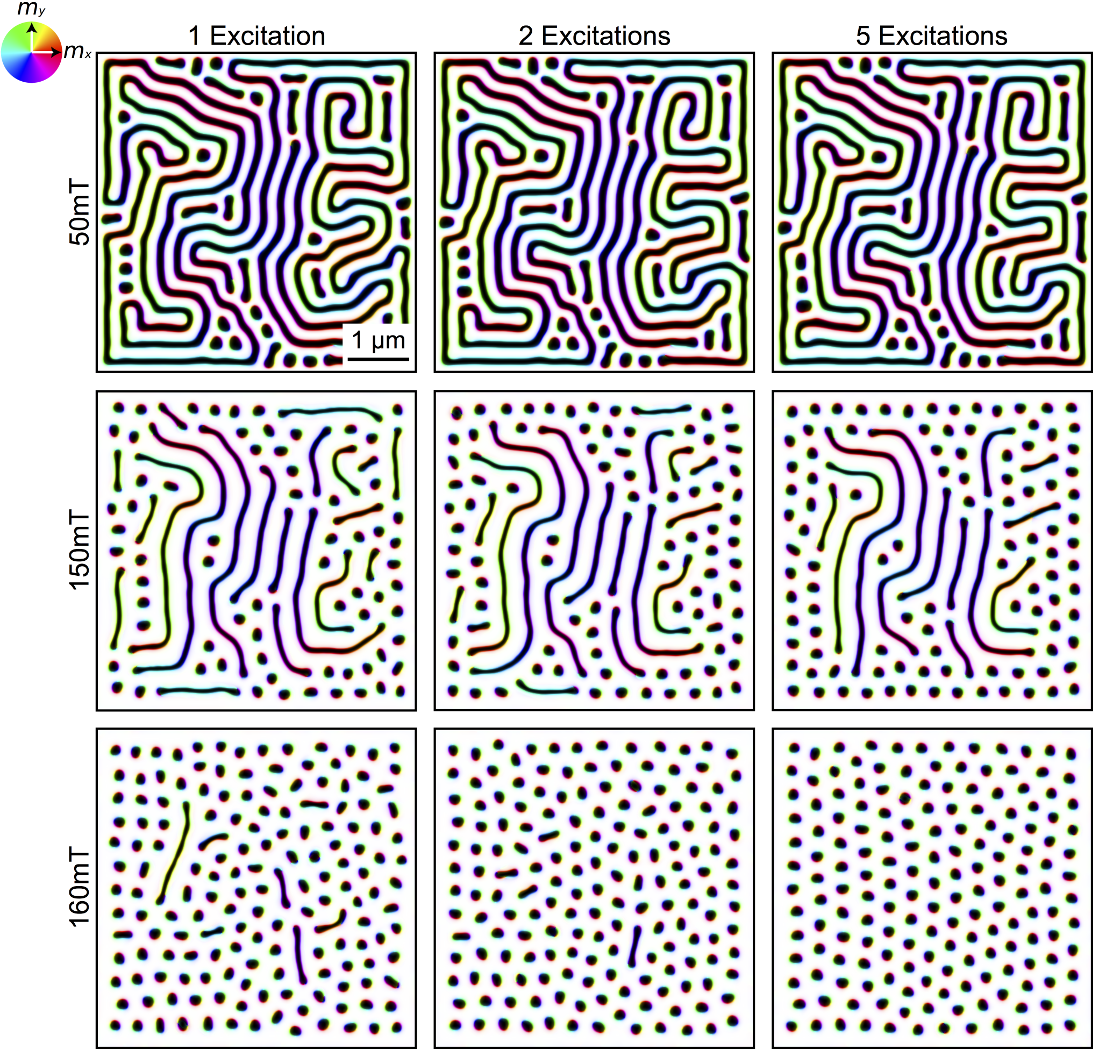}
     \caption{\textbf{Skyrmion nucleation in the micromagnetic simulations.} Spin texture obtained from micromagnetic simulations after laser excitation for different magnetic field regimes. For high magnetic fields ($>\SI{150}{\milli \tesla}$) we observe the transformation of stripe domains into skyrmions upon subsequent excitation, which is absent in case of low magnetic fields (\SI{50}{\milli \tesla}).}
     \label{fig:SkNucSim}
\end{figure}

In Fig.~\ref{fig:SkNucSim} the simulated spin texture is depicted for three magnetic field regimes considering multiple subsequent excitations with an $M_S$ reduction amplitude of \SI{25}{\percent}.

For low magnetic fields (here 50 mT), the spin texture, predominantly consisting out of stripe domains, is stable, i.e., it remains unaffected by laser excitation. Increasing the magnetic field to 150 mT transforms a part of the stripe domains into skyrmions, leading to a mixed phase ground state. Subsequent excitations transform additional stripe domains to skyrmions. A stable equilibrium between stripe domains and skyrmions emerges after few excitation cycles. Thus, a metastable state is created during the very first laser excitations. As a consequence, it is not possible to capture this process with our stroboscopic pump-probe measurement scheme. For slightly higher magnetic fields (\SI{160}{\milli \tesla}) the amount of stripe domains in the ground state is further reduced, facilitating complete transformation of the remaining stripes into skyrmions upon additional excitation.

\begin{figure}[htb]
     \centering
     \includegraphics[width=0.8\columnwidth]{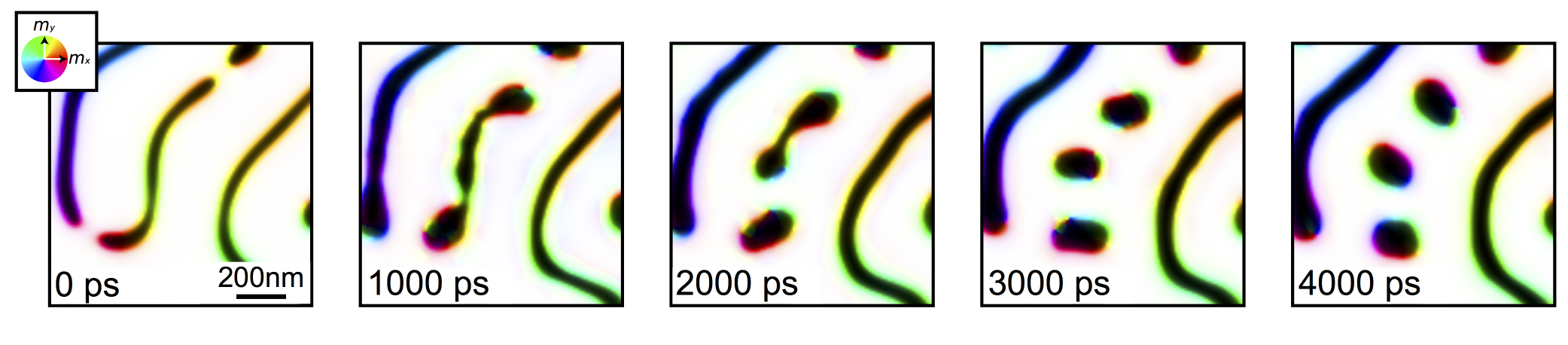}
     \caption{\textbf{Temporal evolution of stripe to skyrmion transformation.} Time evolution of skyrmion nucleation from stripe domains from micromagnetic simulations. We observe the transformation of stripe domains into skyrmions upon subsequent excitation for high magnetic fields ($>\SI{150}{\milli \tesla}$).}
     \label{fig:StSkTrafo}
\end{figure}

We find that this kind of skyrmion nucleation is mediated by the induced motion of the stripe domains and therefore strongly depends on the excitation amplitude. The process of skyrmion nucleation from a stripe domain is depicted in Fig.~\ref{fig:StSkTrafo}. Note that no skyrmion nucleation is observed in micromagnetic simulation upon excitation with an $M_S$ reduction amplitude of \SI{5}{\percent}.

\subsubsection{Fluence-dependence of B/SK nucleation from LTEM studies}

\begin{figure}[htb]
     \centering
     \includegraphics[width=\columnwidth]{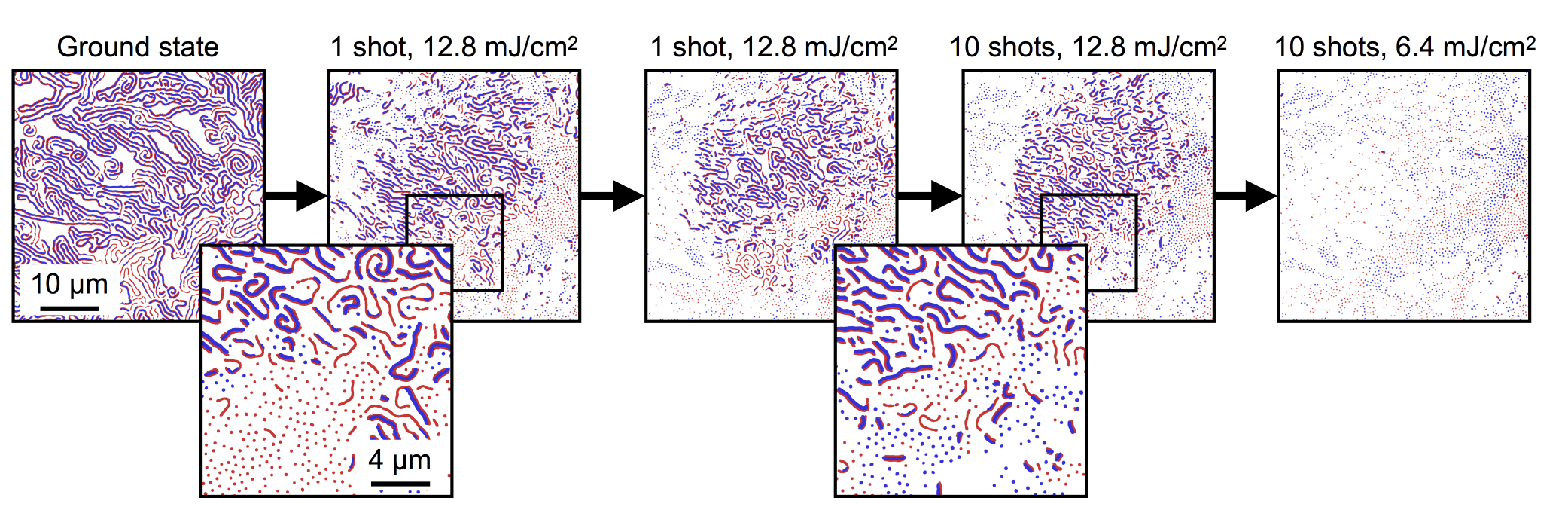}
     \caption{\textbf{Fluence-dependence of spin texture transformation.} Spin texture in dependence of the number of laser pulses. During a downsweep of the magnetic field, the field was set to \SI{195}{\milli \tesla}. Note that the LTEM images were significantly post-processed to enhance the visibility of magnetic skyrmions and their chirality. For details on post-processing see the description of Fig.~\ref{fig:SkfromSt}.}
     \label{fig:Fluence}
\end{figure}

Multiple laser excitation of a pristine stripe phase imaged by LTEM is shown in Fig.~\ref{fig:Fluence}. Upon the first excitation with a fluence of \SI{12.8}{\milli \joule \per \centi \meter \squared} the stripe phase is transformed. Interestingly, a mixed phase appears in the region of highest excitation, while a pure skyrmion phase emerges in its surrounding. Further excitation with the same pulse energy does not macroscopically change the spin texture. However, we observe that the chirality of the skyrmions at the edge of the mixed and the pure skyrmion phase can be inverted by subsequent laser excitation (see offset image zooms). For subsequent weaker excitation, the central mixed phase is transformed into a pure B/SK lattice phase as well. We conclude that the laser pulse has to induce a specific amount of demagnetization in order to nucleate the B/SK lattice phase thereafter. Above a certain threshold of pulse energy this process gets more improbable. Our explanation is that due to the strong demagnetization both the spin texture and the underlying chirality completely vanish. As a result, during the recovery process, the formation of skyrmions is suppressed and single skyrmions only nucleate if by chance the local spin alignment is correct.

\end{document}